\documentclass[twocolumn,twocolappendix,numberedappendix]{openjournal}

\usepackage{xcolor}
\usepackage{textgreek}
\usepackage[utf8]{inputenc}
\usepackage[english]{babel}

\usepackage{hyperref}
\hypersetup{
    unicode,
    colorlinks=true,
    linkcolor=linkcolor,
    citecolor=linkcolor,
    filecolor=linkcolor,
    urlcolor=linkcolor,
}
\usepackage{color,colortbl}
\definecolor{linkcolor}{rgb}{0.0,0.3,0.5}
\usepackage{tensind}
\tensordelimiter{?}
\DeclareGraphicsExtensions{.bmp,.png,.jpg,.pdf}
\usepackage{verbatim}
\usepackage[normalem]{ulem}
\usepackage{orcidlink}
\usepackage{enumitem}
\usepackage{booktabs}
\usepackage{ragged2e}

\graphicspath{ {./figures/} }

\usepackage{ifthen}
\newcommand{\safeincludegraphics}[2][]{%
  \includegraphics[#1]{#2}%
}

\usepackage[T1]{fontenc}

\DeclareRobustCommand{\VAN}[3]{#2}
\let\VANthebibliography\thebibliography
\def\thebibliography{\DeclareRobustCommand{\VAN}[3]{##3}\VANthebibliography}

\usepackage{graphicx}	
\usepackage{adjustbox}  
\usepackage{amsmath}	
\usepackage{wasysym}    
\usepackage{bm}         
\usepackage{xspace}
\usepackage{multirow}
\makeatletter 
  \patchcmd{\NAT@citex}
    {\@citea\NAT@hyper@{%
      \NAT@nmfmt{\NAT@nm}%
      \hyper@natlinkbreak{\NAT@aysep\NAT@spacechar}{\@citeb\@extra@b@citeb}%
      \NAT@date}}
    {\@citea\NAT@nmfmt{\NAT@nm}%
    \NAT@aysep\NAT@spacechar\NAT@hyper@{\NAT@date}}{}{}

  \patchcmd{\NAT@citex}
    {\@citea\NAT@hyper@{%
      \NAT@nmfmt{\NAT@nm}%
      \hyper@natlinkbreak{\NAT@spacechar\NAT@@open\if*#1*\else#1\NAT@spacechar\fi}%
        {\@citeb\@extra@b@citeb}%
      \NAT@date}}
    {\@citea\NAT@nmfmt{\NAT@nm}%
    \NAT@spacechar\NAT@@open\if*#1*\else#1\NAT@spacechar\fi\NAT@hyper@{\NAT@date}}
    {}{}
\makeatother \usepackage{amssymb}

\newtoggle{referee}            %
\togglefalse{referee}          %
\iftoggle{referee}{
    \usepackage{color,soul}    %
}{
}                              %

\newcommand\Msun{\text{M}_{\astrosun}}

\DeclareRobustCommand{\HI}{{\text{H}\,\textsc{i}}\xspace} 
\DeclareRobustCommand{\HII}{{\text{H}\,\textsc{ii}}\xspace} 
\DeclareRobustCommand{\HeI}{{\text{He}\,\textsc{i}}\xspace} 
\DeclareRobustCommand{\HeII}{{\text{He}\,\textsc{ii}}\xspace} 
\DeclareRobustCommand{\thesan}{\mbox{\textsc{thesan}}\xspace}
\DeclareRobustCommand{\thesanzoom}{\mbox{\textsc{thesan-zoom}}\xspace}
\DeclareRobustCommand{\lumina}{\mbox{\textsc{lumina}}\xspace}
\DeclareRobustCommand{\thesanone}{\mbox{\textsc{thesan-1}}\xspace}
\DeclareRobustCommand{\thesantwo}{\mbox{\textsc{thesan-2}}\xspace}

\DeclareRobustCommand{\arepo}{\mbox{\textsc{arepo}}\xspace}
\DeclareRobustCommand{\areport}{\mbox{\textsc{arepo-rt}}\xspace}

\newcommand{\D}{\Delta}
\newcommand{\Tlow}{T_\text{low}}
\newcommand{\Thigh}{T_\text{high}}
\newcommand\orcid[1]{\href{http://orcid.org/#1}{\adjustbox{trim={-.15\width 0 -.15\width 0\height},clip}{\includegraphics[height=9pt]{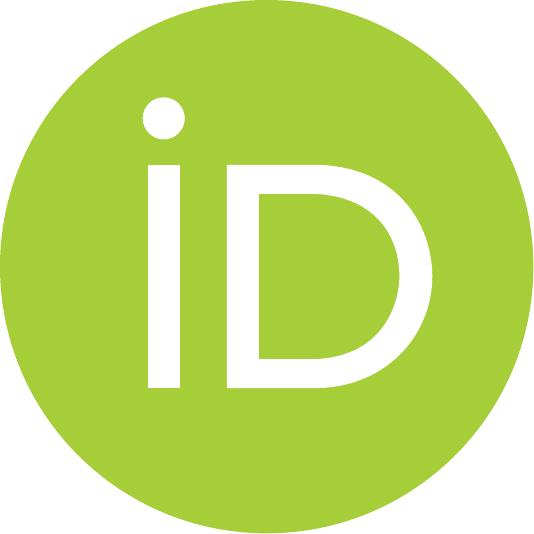}}}}
\newcommand{\appropto}{\mathrel{\vcenter{
  \offinterlineskip\halign{\hfil$##$\cr
    \propto\cr\noalign{\kern2pt}\sim\cr\noalign{\kern-2pt}}}}}

\begin{document}
\title{The Lumina Project: Intergalactic Clumping and Recombination Sinks}

\author{Ali~Sadain\orcid{0009-0009-5786-7469}$^{1\,\star}$}
\email[$^\star$ E-mail: \href{mailto:ali.sadain@utdallas.edu}{ali.sadain@utdallas.edu}]{}
\author{Aaron~Smith\orcid{0000-0002-2838-9033}$^{1}$}
\author{Oliver~Zier\orcid{0000-0003-1811-8915}$^{2}$}
\author{Hyunbae~Park\orcid{0000-0002-7464-7857}$^{3}$}
\author{Teodora-Elena~Bulichi\orcid{0000-0001-8174-6389}$^{4}$}
\author{Xuejian~Shen\orcid{0000-0002-6196-823X}$^{2,4}$}
\author{Rahul~Kannan\orcid{0000-0001-6092-2187}$^{5}$}
\author{Rongrong~Liu\orcid{0000-0003-0685-3525}$^{2}$}
\author{Yuan~Bian\orcid{0000-0002-0210-946X}$^{1}$}
\author{Volker~Springel\orcid{0000-0001-5976-4599}$^{6}$}
\author{Mark~Vogelsberger\orcid{0000-0001-8593-7692}$^{4,7}$}
\author{Sonja~M.~Koehler\orcid{0009-0008-0814-3328}$^{2}$}
\author{Lars~Hernquist\orcid{0000-0001-6950-1629}$^{2}$}
\affiliation{$^{1}$Department of Physics, The University of Texas at Dallas, Richardson, Texas 75080, USA}
\affiliation{$^{2}$Center for Astrophysics $|$ Harvard $\&$ Smithsonian, 60 Garden Street, Cambridge, MA 02138, USA}
\affiliation{$^{3}$Center for Theoretical Physics of the Universe, Institute for Basic Science, Daejeon, 34126, Republic of Korea}
\affiliation{$^{4}$Department of Physics, Kavli Institute for Astrophysics and Space Research, Massachusetts Institute of Technology, Cambridge, MA 02139, USA}
\affiliation{$^{5}$Department of Physics and Astronomy, York University, 4700 Keele Street, Toronto, ON M3J 1P3, Canada}
\affiliation{$^{6}$Max Planck Institute for Astrophysics, Karl-Schwarzschild-Str.\ 1, D-85741 Garching, Germany}
\affiliation{$^{7}$Fachbereich Physik, Philipps Universit\"at Marburg, D-35032 Marburg, Germany}

\begin{abstract}
Recombinations during the Epoch of Reionization are intrinsically inhomogeneous, with different regions of the intergalactic medium contributing unevenly depending on their density, temperature, ionization state, and spatial patchiness. We combine the high- and medium-resolution $95.5\,\mathrm{cMpc}$ \thesanone and \thesantwo runs with the significantly larger $500\,\mathrm{cMpc}$ \lumina simulation to measure clumping factors and recombination rates consistently across different resolutions and box sizes. We consider the standard $\Delta<100$ ionized hydrogen clumping factor, $\mathcal{C}_\text{\HII} \equiv \langle n_\text{\HII}^2\rangle_{<100}/\langle n_\text{\HII}\rangle^2$, and a recombination-weighted clumping factor, $\mathcal{C}_\text{rec}$. Despite differences in resolution, volume, and reionization history, the simulations show an approximately universal clumping evolution at the $\sim10$--$20\%$ level when parametrized by the global ionized fraction $x_\text{\HII}$ rather than by redshift. Across all simulations, $\mathcal{C}_\text{rec}$ remains systematically below $\mathcal{C}_\text{\HII}$, with the discrepancy increasing toward lower redshift as photoheating suppresses recombinations. In \lumina, the density-only prescription overpredicts the instantaneous recombination rate by factors of $1.34$ at $z\simeq8$ and $2.02$ at $z\simeq5$, and the cumulative recombination count by a factor of $1.53$ by $z\simeq5$. Mapping the recombination budget in the joint overdensity--temperature plane reveals that the dominant recombination ridges closely follow simple analytic thermal equilibrium bands derived from the Hui--Gnedin equation of state. Finally, we introduce a phase-space recombination integral and define a phase-space clumping factor, $\mathcal{C}_{\rm ps}(\Delta,T)$, which isolates the intrinsic recombination enhancement associated with ionization structure and thermal state at fixed overdensity and temperature while closely reproducing the globally measured recombination rate obtained from direct cell-by-cell calculations. This formulation naturally decomposes the total recombination budget as a function of overdensity, temperature, and ionization state.
\end{abstract}


\maketitle


\section{Introduction}

Cosmic reionization marks the transition of the intergalactic medium (IGM) from a predominantly neutral state to a fully ionized state. Driven by ultraviolet photons produced by galaxies, this process unfolded over several hundred million years, beginning as the earliest luminous sources formed at $z\gtrsim15$ and ending once ionized regions percolated throughout most of the Universe by $z \lesssim 6$. On global scales, reionization is governed by the competition between sources of ionizing photons and the sinks that consume them through recombinations. In the simplest photon-counting picture, each hydrogen atom must be ionized at least once, but the same atom can recombine and be ionized repeatedly. Accurately accounting for these repeated recombinations is therefore essential for closing the ionizing photon budget and interpreting the timing, duration, and morphology of reionization \citep{BarkanaLoeb2001,Madau1999,Robertson2015,Shull2012,Gnedin2022Review}. In homogeneous models, the recombination rate is usually boosted by a single clumping factor that summarizes the effect of unresolved density structure \citep{Madau1999,Furlanetto2004,Pawlik2009,Finlator2012,Chen2020SCORCH}. This approach is simple and useful, but marginalizes over several distinct physical effects, including cosmological large-scale structure, ionization patchiness, gas temperature, self-shielding, and the spatial distribution of absorbers.

A long line of work has shown that the clumping factor is not a universal constant, but depends on how the IGM is defined, which gas phases are included, the resolution of the calculation, and the ionization and thermal history of the gas \citep{Pawlik2009,RaicevicTheuns2011,Finlator2012,JeesonDaniel2014,KaurovGnedin2015,Chen2020SCORCH}. Photoheating suppresses recombinations both by raising the gas temperature and thereby lowering the recombination coefficient $\alpha(T)$, as described by \citet[][henceforth HG97]{HuiGnedin1997}, and by pressure-smoothing small-scale density structure \citep{Pawlik2009,McQuinn2016}. Dense gas in filaments, halo outskirts, and self-shielded systems, on the other hand, can remain efficient at consuming ionizing photons and setting the ionizing mean free path \citep{MiraldaEscude2000,McQuinn2011,Rahmati2013,DaviesFurlanetto2016, park2016, Daloisio2020}. These effects make recombinations strongly inhomogeneous and introduce significant uncertainty into semi-analytic and semi-numerical reionization models \citep{Mao2020,Bianco2021,Chen2020SCORCH}.

The importance of recombination sinks has also been emphasized in simulations that model reionization morphology and the growth of ionized bubbles \citep{FurlanettoOh2005,Sobacchi2014, Neyer2024}. Analytic excursion-set models and large-scale radiative-transfer calculations show that the topology of ionized regions is shaped not only by the abundance and clustering of sources, but also by small-scale absorbers, recombinations, and the evolving mean free path \citep{Furlanetto2004,Iliev2006,McQuinn2007,Mesinger2011,Cain2022b}. Subgrid treatments of recombinations are therefore common in large-volume calculations, where the relevant absorbing systems are not always resolved \citep{Mao2020,Bianco2021,Chen2020SCORCH}. High-resolution studies of the small-scale IGM show that these absorbers are dynamic, with recently ionized filaments and minihalo gas temporarily boosting recombinations and opacity before photoheating and pressure smoothing erase these structures \citep{Daloisio2020,Nasir2021,Cain2024a,Cain2026SAGUARO}. A central challenge is to build recombination prescriptions that are simple enough for reduced models yet faithful to the density, temperature, and ionization structure seen in radiation-hydrodynamic simulations.

Observational inferences of the ionizing background, ionizing emissivity, Ly$\alpha$ opacity, and mean free path near the end of reionization are also tied to this problem. The conversion from Ly$\alpha$ forest opacity to a photoionization rate depends on the density and temperature structure of the gas, while the inferred emissivity depends on the abundance of absorbers and the recombination rate in the ionized IGM \citep{BoltonHaehnelt2007,BeckerBolton2013,HaardtMadau2012,Puchwein2019,FaucherGiguere2020}. Recent direct and forward-modeled constraints find rapid mean-free-path evolution over $5\lesssim z\lesssim6$ \citep{Becker2021MFP,Gaikwad2023MFP,Zhu2023MFP,Davies2024MFP}. The rapid evolution of the mean free path and Ly$\alpha$ opacity at $z\sim5$--6 has motivated models with strong spatial fluctuations in the ionizing background and a significant role for clustered absorbers \citep{DaviesFurlanetto2016,McQuinn2011,Cain2021,lewis2022short,croc120cMpc, Park2024}. Alternative explanations invoke residual temperature fluctuations from an extended and inhomogeneous reionization process or persistent neutral islands near the end of reionization, highlighting the close connection between Ly$\alpha$ forest observables, the thermal state of the IGM, and the abundance of absorbers \citep{Daloisio2015,Kulkarni2019}. A robust treatment of recombination sinks is therefore required both for source-counting arguments and for interpreting post-overlap IGM constraints.

Recent work has shown that this problem is highly definition-dependent. Effective clumping factors inferred from the ionizing mean free path and photoionization rate near the end of reionization do not necessarily coincide with the closure terms measured in simulations. The inferred value depends on how the ionized IGM is separated from galaxies and self-shielded absorbers, whether averages are volume- or mass-weighted, and whether temperature and ionization structure are explicitly included \citep{KaurovGnedin2015,Chen2020SCORCH,Davies2024}. This distinction is important as short mean-free-path measurements have reignited debate over the magnitude of sink-driven photon losses during the final stages of reionization. In this paper, we therefore treat clumping factors as reduced descriptions of the recombination budget whose meaning depends on the adopted gas selection and weighting.

Modern radiation-hydrodynamic simulations now follow the coupled evolution of galaxies, radiation, gas dynamics, and non-equilibrium thermochemistry in a self-consistent way. Building on earlier large-scale radiative-transfer calculations \citep{Iliev2006,TracCenLoeb2008,AubertTeyssier2010}, recent simulation suites include CROC \citep{Gnedin2014design,croc120cMpc}, CoDa \citep{ocvirk2016cosmic,ocvirk2020cosmic,lewis2022short,Ocvirk2025}, Aurora \citep{aurora2017}, SPHINX \citep{rosdahl2018sphinx,sphinx2022}, SCORCH \citep{Doussot2019SCORCH}, SPICE \citep{bhagwat2024spice}, \thesan \citep{Kannan2022a,Smith2022,Garaldi2022,Garaldi2024}, and \lumina \citep{Zier2026}. The semi-numerical AMBER model provides a complementary large-volume framework \citep{Trac_2022}. These simulations have made it possible to connect the global reionization history to the internal structure of the IGM and circumgalactic medium (CGM), including the distribution of absorbers, the topology of ionized regions, and the thermal imprint of reionization.

In this work, we focus on the \thesan and \lumina simulations. The \thesan suite combines the IllustrisTNG galaxy-formation model with on-the-fly radiative transfer and non-equilibrium thermochemistry, and has been used to study, e.g., reionization histories, high-redshift galaxies, ionizing escape fractions, Ly$\alpha$ transmission, bubble size distributions, and line-intensity mapping \citep{Kannan2022b,Yeh2023,Xu2023,Neyer2025,Almualla2025,Zhao2026}, which all reveal different aspects of galaxy--IGM interactions. The new \lumina simulation extends this framework to a much larger volume with updated radiation and galaxy-formation physics \citep{Zier2026}. It has already been used to study the high-redshift active galactic nucleus (AGN) population and its large-volume hydrogen- and helium-reionization histories, including their implications for the cosmic microwave background (CMB) Thomson optical depth \citep{Shen2026, Smith2026}. Together, \thesanone, \thesantwo, and \lumina provide a useful hierarchy for this study. Specifically, \thesanone resolves small-scale gas structure in a $95.5\,\mathrm{cMpc}$ box, \thesantwo provides a direct lower-resolution comparison using the same initial conditions, and \lumina maintains a resolution comparable to \thesantwo over a large $500\,\mathrm{cMpc}$ cosmological volume. This allows us to ask how recombination statistics depend on resolution, volume, and reionization history.

The thermal state of the IGM is central to this question. In the low-density photoionized IGM, gas evolves toward a temperature--density relation set by the balance between photoheating, radiative cooling, and cosmological expansion described by HG97 and subsequent work \citep{Theuns1998,Schaye1999,Ricotti2000,McQuinn2016}. The thermal history retains memory of when and how gas was reionized, because recently ionized regions can remain hotter than regions ionized earlier \citep{TracCenLoeb2008,McQuinnUptonSanderbeck2016,Daloisio2018}. Recent numerical work has emphasized that the emergence of the temperature--density relation during reionization is closely tied to the global ionization state of the IGM \citep{Wells2024}. Since the hydrogen recombination coefficient depends strongly on temperature, the recombination budget must be understood in the joint space of density, temperature, and ionization state.

Motivated by these developments, we analyze hydrogen recombinations in the \thesan and \lumina simulations using both global and phase-resolved diagnostics. We first compare standard density-based clumping factors with recombination-weighted clumping factors that include the temperature dependence of the recombination coefficient. We then map recombinations into $(\Delta,T)$ phase space, where $\Delta\equiv\rho/\bar{\rho}$ is the overdensity, and compare the dominant recombination regions to simple analytic equilibrium curves. Finally, we introduce a phase-space clumping factor that isolates the intrinsic recombination boost due to ionization patchiness and temperature at fixed $(\Delta,T)$, separating this local enhancement from the volume, density, and ionized-fraction weights that determine its global importance.

Our goal is to give a physical accounting of recombination sinks beyond universal clumping factors. We consider the following: How much error is introduced when a density-only clumping factor ignores the temperature dependence of recombinations? Does clumping evolve more clearly with redshift, or with the volume-weighted global ionized fraction $x_{\HII}$? Which regions of $(\Delta,T)$ phase space dominate the recombination budget, and how do they relate to simple thermal-equilibrium expectations? How spatially concentrated are the dominant recombination sinks? Can a phase-space clumping factor separate the extra boost from ionization and temperature structure from the simpler effect that dense gas occupies little volume?

The paper is organized as follows. In Section~\ref{sec:methods} we describe the simulations and define the global, spatial, and phase-space recombination quantities used in the analysis. In Section~\ref{sec:results} we present the evolution of global clumping factors, the spatial concentration of recombinations, and the phase-space structure of recombination sinks. In Section~\ref{sec:discussion} we discuss the implications of these results, and in Section~\ref{sec:conclusions} we summarize the main conclusions. Appendix~\ref{app:thermal_bands} derives the analytic thermal-equilibrium bands used in the phase-space analysis. Appendices~\ref{app:threshold_sensitivity} and \ref{app:fits_cxhii} examine the overdensity-threshold sensitivity and provide fitting functions for the clumping factors, respectively.

\section{Simulation and methods}
\label{sec:methods}

\subsection{Simulation overview}

We analyze three fully coupled radiation-hydrodynamic simulations that address the main numerical trade-offs relevant for recombination sinks during reionization: the fiducial \thesanone and lower-resolution \thesantwo runs from the \thesan suite \citep{Kannan2022a,Smith2022,Garaldi2022,Garaldi2024}, and the much larger-volume \lumina simulation \citep{Zier2026}. All three calculations follow the coupled evolution of dark matter, gas, stars, and ionizing radiation with on-the-fly radiative transfer and non-equilibrium thermochemistry, but they differ in cosmological volume, mass resolution, and source modeling. \thesanone is the highest resolution run, resolving small-scale gas structure and the internal regions of halos down to the atomic cooling limit in a $(95.5\,\mathrm{cMpc})^3$ box with $2\times2100^3$ resolution elements and baryonic mass resolution $m_b = 5.8\times10^5\,\Msun$. \thesantwo adopts the exact same initial conditions but with $2\times$ ($8\times$) coarser spatial (mass) resolution, providing a direct convergence study of the original \thesan model. On the other hand, \lumina includes additional high-energy radiation sources from accreting black holes, high-mass X-ray binaries, and hot ISM gas, follows both hydrogen and helium reionization, and incorporates numerical improvements for large-scale radiation-hydrodynamic simulations. It also offers a significantly larger $(500\,\mathrm{cMpc})^3$ volume with $2\times6000^3$ resolution elements and baryonic mass resolution $m_b = 3.6\times10^6\,\Msun$, corresponding to $1.84\times$ ($6.2\times$) lower spatial (baryonic mass) resolution than \thesanone but 143 times larger volume. Prioritizing volume probes a wider range of IGM environments for greater statistical power. This hierarchy allows us to assess how recombination statistics depend on numerical resolution, cosmological volume, and reionization history. Table~\ref{tab:simulation_properties} summarizes the basic numerical properties of the runs.

All simulations assume a flat $\Lambda$CDM cosmology with a minor update to use parameters from \citet{Planck2020} in \lumina compared to \cite{Planck2015_cosmo} in \thesan. For the detailed generation of the initial conditions, escape-fraction calibrations, and source prescriptions, we refer the reader to the presentation papers for \thesan and \lumina \citep{Kannan2022a,Garaldi2024,Zier2026}.

\subsection{Radiation hydrodynamics and galaxy formation}

The simulations are run with the moving-mesh code \arepo \citep{springel2010pur,pakmor2016improving,weinberger2020arepo}, which solves the Euler equations on an unstructured Voronoi mesh with adaptive spatial resolution and quasi-Lagrangian mesh motion, coupled to self-gravity. Ionizing radiation is evolved on the same mesh with \areport, a moment-based radiative-transfer solver using the M1 closure, coupled to non-equilibrium hydrogen and helium chemistry \citep{Kannan2019,Levermore1984,Dubroca1999}. Both simulations adopt a reduced speed of light approximation with $\tilde{c}=0.2\,c$ \citep{Gnedin2001}. The \lumina implementation additionally incorporates GPU acceleration and further large-volume scalability improvements \citep{Gadget4,Pakmor2023,zier2024adapting,Zier2026}. Galaxy formation in both suites follows the IllustrisTNG framework \citep{Vogelsberger2013,IllustrisIntro,IllustrisNature,Weinberger2017,Pillepich2018}. In brief, star formation proceeds stochastically in cold, dense gas treated with an effective two-phase interstellar medium (ISM) equation of state \citep{Springel2003}, stellar feedback is implemented with a kinetic-wind model, and black holes are seeded in sufficiently massive halos and evolve through accretion, mergers, and feedback \citep{Weinberger2017,Pillepich2018}. \lumina incorporates several updates to the black-hole treatment relative to \thesan, including corrections to the black-hole accretion and feedback implementation and a self-consistent treatment of AGN as ionizing radiation sources \citep{Bulichi2025,Zier2026}. These modifications are more consistent with the original IllustrisTNG black-hole model while simultaneously following AGN-driven \HeII reionization.

While \thesan is evolved to $z = 5.5$, \lumina extends down to $z = 3$, split into hydrogen and helium reionization stages. Specifically, at $z>4.75$ the radiation field is discretized into six energy bins which include the \HI, \HeI, and \HeII ionization thresholds together with three higher-energy X-ray bins. The dominant ionizing sources are stellar populations based on \textsc{bpass} v2.2.1 spectra \citep{BPASS2017,Stanway2018} and a Chabrier initial mass function \citep[IMF;][]{Chabrier2003}, with constant stellar escape fractions of $f_{\text{esc},\star} = 0.37$ for \thesanone and $f_{\text{esc},\star} = 0.18$ for \lumina, calibrated to reproduce the desired hydrogen-reionization history. The values differ mainly because \lumina treats star-forming cells as transparent for numerical stability and other modeling choices. \lumina also includes hard radiation from accreting black holes and X-ray emission associated with high-mass X-ray binaries and hot ISM gas \citep{mcquinn2009he,Shen2020,2013Fragos-HMXB,2016Fragos-Erratum,Madau2017,2012bMineo-HotISM,2014Pacucci}. These high-energy components are potentially relevant for the present paper because they can pre-heat gas well ahead of the main ultraviolet ionization fronts \citep{Pritchard2007,ma2018,2018Eide,2020Eide}. At $z=4.75$, once hydrogen reionization is effectively complete, photons below the \HeII edge are replaced by a spatially uniform metagalactic background and the remaining transported photon bins are merged to optimize the subsequent \HeII calculation.

\begin{table}
\centering
\caption{\justifying \noindent \textup{Simulation properties for the \lumina \citep{Zier2026} and \thesan \citep{Kannan2022a, Garaldi2024} runs used in this work. Columns list the comoving box size ($L_\text{box}$), the initial gas and dark-matter resolution-element counts ($N_{\rm part}$), the baryonic and dark-matter mass resolutions ($m_b$, $m_{\rm dm}$), the comoving Plummer-equivalent dark-matter softening and the minimum gas softening ($\epsilon_{\rm dm}$, $\epsilon_{\rm gas, min}$), and the final redshift ($z_{\rm final}$).}}
\label{tab:simulation_properties}
\footnotesize
\setlength{\tabcolsep}{2.8pt}
\renewcommand{\arraystretch}{1.1}
\begin{tabular}{lcccccc}
\toprule
Run & $L_\text{box}$ & $N_{\rm part}$ & $m_b$ & $m_{\rm dm}$ & $\epsilon$ & $z_{\rm final}$ \\
& ($\mathrm{cMpc}$) & & (${\rm M_\odot}$) & (${\rm M_\odot}$) & ($\mathrm{ckpc}$) & \\
\midrule
\lumina & 500 & $6000^3$ & $3.6\times10^6$ & $1.9\times10^7$ & 1.77/0.44 & 3.0 \\
\thesanone & 95.5 & $2100^3$ & $5.8\times10^5$ & $3.1\times10^6$ & 2.2/2.2 & 5.5 \\
\thesantwo & 95.5 & $1050^3$ & $4.7\times10^6$ & $2.5\times10^7$ & 4.4/4.4 & 5.5 \\
\bottomrule
\end{tabular}
\renewcommand{\arraystretch}{1}
\end{table}

\subsection{Analysis fields and IGM gas selection}

All quantitative statistics in this paper are computed directly from the native Voronoi gas cells. For each snapshot we use the following quantities for recombination calculations: the mass density $\rho$, hydrogen number density $n_\text{H} \equiv X\,\rho / m_\text{H}$ with hydrogen mass fraction $X = 0.76$ and atomic mass $m_\text{H}$, ionized hydrogen fraction $x_{\HII} \equiv n_{\HII} / n_\text{H}$, electron abundance $x_e \equiv n_e / n_\text{H}$, temperature $T$, and cell volume $V$.

When computing global statistics we work in physical units and adopt the conventional overdensity cut $\Delta \equiv \rho / \bar{\rho} < 100$ in the \emph{numerators} of the IGM recombination and clumping measures. This excludes the densest ISM/CGM gas within galaxies while retaining the diffuse IGM/CGM components that act as large-scale photon sinks. The threshold removes only a small fraction of the simulation volume, with more than $99.98\%$ of the volume satisfying $\Delta<100$ at all redshifts considered. The normalization densities in the denominators, however, are computed over the entire gas distribution. This convention makes the resulting factors directly applicable to global photon-counting equations in which the clumping correction multiplies the square of a cosmic mean density.

Throughout this paper, angle brackets without a subscript denote a volume average over all gas,
\begin{equation}
  \langle f\rangle \equiv
  \frac{\sum_i f_i V_i}{\sum_i V_i} \, ,
  \label{eq:all_gas_average}
\end{equation}
whereas a subscript $<100$ restricts the contributing cells to gas with overdensity $\Delta<100$,
\begin{equation}
  \langle f\rangle_{<100} \equiv
  \frac{\sum_{i:\,\Delta_i<100} f_i V_i}{\sum_i V_i} \, .
  \label{eq:selected_gas_moment}
\end{equation}
Thus, the overdensity cut restricts which cells contribute to a numerator, but it does not renormalize that numerator by the smaller selected volume. All moments below are evaluated as explicit cell-wise sums using these conventions. Sensitivity to the numerator threshold is discussed in Appendix~\ref{app:threshold_sensitivity}.

\begin{figure*}
    \centering
\safeincludegraphics[width=\textwidth]{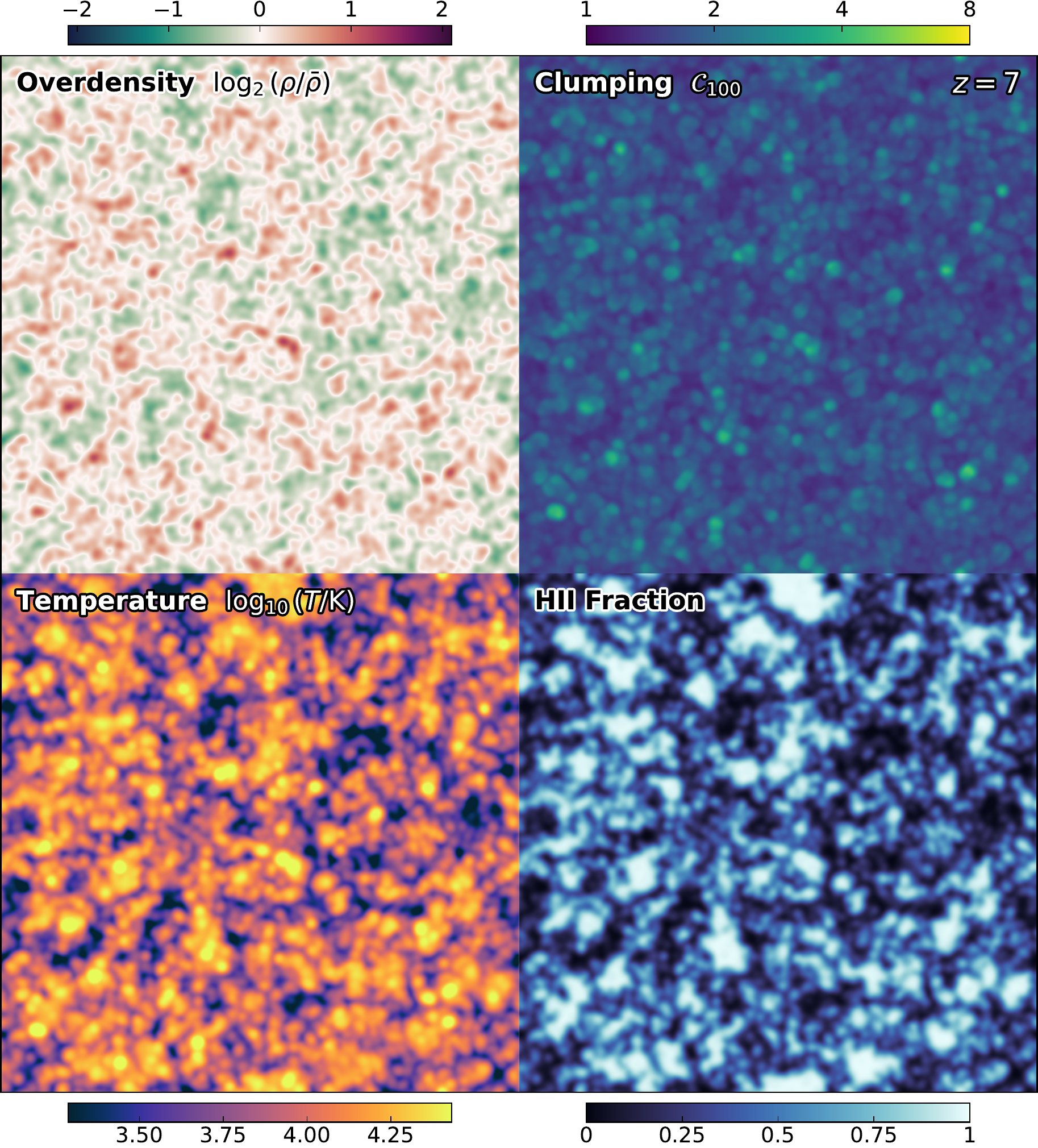}
\caption{
Real-space view of the intergalactic medium at $z=7$ in \lumina, showing overdensity, clumping factor $\mathcal{C}_{100}$ defined in Eq.~\eqref{C_100 equation} evaluated within a local smoothing volume, temperature, and ionized fraction $x_{\mathrm{HII}}$. The maps are projected through the same slice and then smoothed on a common $3.125\,\mathrm{cMpc}$ scale. Colors show raw projected field values with each panel independently normalized to display its dynamic range. Dense regions correspond to enhanced clumping, while the temperature map reveals the $\sim10^4\,\mathrm{K}$ photoheated phase, and the ionization field traces the inhomogeneous progression of reionization. This visualization illustrates the spatial origin of recombination sinks and their connection to the underlying density and thermal structure, which are analyzed quantitatively in the following sections.
}
    \label{fig:realspace_fields}
\end{figure*}

In addition to the cell-based measurements, we evaluate a local clumping factor by averaging the hydrogen density field within a Gaussian kernel $\mathcal{K}(\bm{r},R)$ centered at position $\bm{r}$ with smoothing scale $R=3.125\,{\rm cMpc}$. Specifically, we define
\begin{equation}
\mathcal{C}_{100}(\bm{r})
\equiv
\frac{\left\langle n_{\rm H}^2 \right\rangle_{\mathcal{K}(\bm{r},R)}}
     {\left\langle n_{\rm H} \right\rangle_{\mathcal{K}(\bm{r},R)}^2} \, ,
     \label{C_100 equation}
\end{equation}
where the averages are computed over all gas within the kernel, restricting to cells with $\Delta<100$. The value $\mathcal{C}_{100}=1$ corresponds to a uniform medium, while inhomogeneous structures yield $\mathcal{C}_{100}>1$. The simulations provide high-cadence Cartesian outputs that are useful for visualizing the large-scale morphology of recombination sinks. These analysis products deposit mass- and volume-weighted fields onto a uniform grid using an adaptive second-order cloud-in-cell scheme. In Figure~\ref{fig:realspace_fields} we use these grids to show real-space images of the gas overdensity, clumping, temperature, and \HII fraction, smoothed with the same Gaussian kernel, highlighting the relative homogeneity of the IGM on large scales.

\subsection{Global clumping factors and recombination rates}

We focus on two global clumping measures. The standard ionized-hydrogen clumping factor is
\begin{equation} \label{eq:CHII_def}
  \mathcal{C}_{\HII}(z) \equiv
  \frac{\langle n_{\HII}^2 \rangle_{<100}}
       {\langle n_{\HII} \rangle^2} \, ,
\end{equation}
which measures the contribution of structure at $\Delta<100$ relative to the square of the all-gas mean ionized-hydrogen density. Because the true recombination rate depends on both density and temperature through the Case~B coefficient $\alpha_\text{B}(T)$, the traditional density-based clumping factor does not directly capture the net rate at which ionizing photons are lost \citep[e.g.][]{Pawlik2009,Finlator2012,KaurovGnedin2015,Sobacchi2014,park2016}. We therefore define the recombination-weighted clumping factor
\begin{equation} \label{eq:Crec_def}
  \mathcal{C}_\text{rec}(z) \equiv
  \frac{\langle \alpha_\text{B}(T)\,n_e\,n_\text{\HII}\rangle_{<100}}
       {\alpha_\text{B}(10^4\,\text{K})\,
        \langle n_e\rangle\,
        \langle n_\text{\HII}\rangle} \, ,
\end{equation}
which measures the selected IGM/CGM contribution relative to a homogeneous all-gas reference model at $10^4$\,K. If $n_e$ is proportional to $n_{\HII}$ and the selected gas is nearly isothermal at $10^4$\,K, $\mathcal{C}_{\rm rec}$ approaches $\mathcal{C}_{\HII}$.

Throughout the recombination-budget analysis we use the Case~B hydrogen recombination coefficient from HG97. Case~B excludes direct recombinations to the ground state, whose emitted ionizing photons are absorbed locally and therefore do not represent a net loss from the ionizing photon budget under the on-the-spot approximation. Using Case~A in the sink term would require the corresponding recombination-radiation contribution to be added explicitly to the source term. The exact selected-gas recombination rate density per total simulation volume is
\begin{equation} \label{eq:Gamma_exact}
  \Gamma_\text{rec}(z) =
  \big\langle \alpha_\text{B}(T)\,n_e\,n_\text{\HII}\big\rangle_{<100} \, ,
\end{equation}
so that by the definition of $\mathcal{C}_\text{rec}$,
\begin{equation} \label{eq:Gamma_in_terms_of_Crec_preChi}
  \Gamma_\text{rec}(z) = \mathcal{C}_\text{rec}(z)\,\alpha_\text{B}(10^4\,\text{K})\,\langle n_e\rangle\,\langle n_\text{\HII}\rangle \, .
\end{equation}
To account for additional electrons contributed by ionized helium we further define
\begin{equation} \label{eq:chie_def}
  \chi_e(z) \equiv \frac{\langle n_e\rangle}{\langle n_\text{\HII}\rangle} \, ,
\end{equation}
which is typically $\chi_e \approx 1.08$ as helium is predominantly singly ionized during hydrogen reionization. Using this definition, Eq.~\eqref{eq:Gamma_in_terms_of_Crec_preChi} becomes
\begin{equation} \label{eq:Gamma_in_terms_of_Crec}
  \Gamma_\text{rec}(z) = \mathcal{C}_\text{rec}(z)\,\alpha_\text{B}(10^4\,\text{K})\,\chi_e(z)\,\langle n_\text{\HII}\rangle^2 \, .
\end{equation}
The density-only comparison shown in Figures~\ref{fig:rec-rate}, \ref{fig:Nrec-compare}, and \ref{fig:photon_budget} is
\begin{equation}
  \Gamma_\text{rec}^{(\mathcal C_{\HII})}(z)
  = \mathcal{C}_{\HII}(z)\,\alpha_\text{B}(10^4\,\text{K})\,
  \langle n_\text{\HII}\rangle^2 \, ,
  \label{eq:Gamma_CHII_prescription}
\end{equation}
which deliberately retains only ionized-hydrogen density structure and a fixed temperature; the full electron abundance enters the direct Case~B rate through Eq.~\eqref{eq:Gamma_in_terms_of_Crec_preChi}.
A useful normalized global diagnostic is the instantaneous number of recombinations per hydrogen atom per Hubble time, denoted by a dotted symbol,
\begin{equation} \label{eq:Nrec_def_final}
  \dot{N}_\text{rec}(z) \equiv \frac{\Gamma_\text{rec}(z)}{\langle n_\text{H}\rangle}\,H^{-1}(z) \, .
\end{equation}
This quantity is an instantaneous recombination rate expressed in units of the Hubble time, while the time-integrated number of recombinations per atom is introduced and plotted in Figure ~\ref{fig:photon_budget}.

To isolate the effective thermal weighting while retaining the simulated electron--ion density correlation, we define
\begin{equation}
  \mathcal C_{e\HII}(z) \equiv
  \frac{\langle n_e n_{\HII}\rangle_{<100}}
       {\langle n_e\rangle\langle n_{\HII}\rangle} \, .
  \label{eq:CeHII_def}
\end{equation}
We then define $T_\text{eff}(z)$ implicitly through
\begin{equation} \label{eq:Teff_def}
  \Gamma_\text{rec}(z) = \mathcal{C}_{e\HII}(z)\,\alpha_\text{B}\!\big(T_\text{eff}(z)\!\big)\,\langle n_e\rangle\,\langle n_\text{\HII}\rangle \, ,
\end{equation}
so that Eqs.~\eqref{eq:Gamma_in_terms_of_Crec_preChi} and \eqref{eq:Teff_def} give
\begin{equation} \label{eq:Teff_ratio_final}
  \frac{\alpha_\text{B}\!\big(T_\text{eff}(z)\!\big)}{\alpha_\text{B}(10^4\,\text{K})} = \frac{\mathcal{C}_\text{rec}(z)}{\mathcal{C}_{e\HII}(z)} \, .
\end{equation}
In general, $\mathcal{C}_\text{rec} < \mathcal{C}_{e\HII}$ corresponds to $T_\text{eff} > 10^4$\,K.

\subsection{Overdensity-based statistics}

Working directly with the recombination rates of individual Voronoi gas cells on the adaptive \arepo mesh, we construct overdensity probability density functions (PDFs) that probe the density distribution down to the native resolution limit of each simulation, without additional smoothing or regridding. The volume-integrated rate in units of $\text{s}^{-1}$ is
\begin{equation}
  \dot{\mathcal N}_{\text{rec},i} \equiv \alpha_\text{B}(T_i)\,n_{e,i}\,n_{\text{\HII},i}\,V_i \, ,
\end{equation}
where $V_i$ is the physical volume of cell $i$. The total rate within the fiducial IGM cut ($\Delta<100$) is
\begin{equation}
  \dot{\mathcal N}_\text{rec}^{\Delta<100} \equiv \sum_{i:\,\Delta<100} \dot{\mathcal N}_{\text{rec},i} \, .
\end{equation}
We define the total probability that recombinations occur within overdensity bins denoted by $\Delta$ as
\begin{equation}
  P_\text{rec}(\Delta) \equiv \frac{\dot{\mathcal N}_\text{rec}^\Delta}{\dot{\mathcal N}_\text{rec}^{\Delta<100}} = \frac{\sum_{i:\,\Delta} \dot{\mathcal N}_{\text{rec},i}}{\sum_{i:\,\Delta<100} \dot{\mathcal N}_{\text{rec},i}} \, ,
\end{equation}
and the total volume fraction over the same overdensity bins is $P_V(\Delta) = \sum_{i:\,\Delta} V_i \,/ \sum_{i:\,\Delta<100} V_i$. Both probabilities are normalized such that $\sum P_\text{rec}(\Delta) = \sum P_V(\Delta) = 1$. The cumulative distribution function (CDF) is
\begin{equation}
  F_\text{rec}(\Delta) \equiv \sum_{\Delta' \le \Delta} P_\text{rec}(\Delta') \, ,
\end{equation}
and similar for $F_V(\Delta)$. The relation $F_{\rm rec}(F_V)$ provides a Lorenz-style diagnostic summarized by the corresponding Gini coefficient, $G \equiv 1 - 2\int_0^1 F_{\rm rec}(F_V)\,{\rm d}F_V$. Together, these quantify how a small volume fraction can dominate the total recombination budget \citep[cf.][]{MiraldaEscude2000, FurlanettoOh2005, Sobacchi2014, DaviesFurlanetto2016}.

\subsection{Phase-space recombination statistics}
In addition to density, temperature provides another important degree of freedom for characterizing recombination-dominated regions. Exploring the joint $\Delta$--$T$ phase space can reveal whether the high-$\Delta$ recombination sinks are further stratified by temperature \citep{Finlator2012,KaurovGnedin2015,Daloisio2018}, and improve the physical interpretation of clumping and recombinations, e.g., whether gas was recently photoheated. We therefore define the recombination-weighted phase-space probability dependence as
\begin{equation}
  P_\text{rec}(\Delta, T) = \frac{\sum_{i:\,\Delta,T} \dot{\mathcal{N}}_{\text{rec},i}}{\dot{\mathcal N}_\text{rec}^{\Delta<100}} \, ,
\end{equation}
where the numerator sums $\dot{\mathcal{N}}_\text{rec}$ over all cells in a joint bin $(\Delta, T)$, and the denominator sums over all cells with $\Delta<100$ regardless of temperature. The volume-weighted phase-space probabilities $P_V(\Delta, T)$ are defined similarly. Prominent recombination sinks within specific regions of the $\Delta$--$T$ space can be visually identified in 2D heatmaps, isolating physically distinct regimes contributing to the total recombination budget, such as cool, dense clumps versus hot, shocked gas.

\subsection{Phase-space clumping factor}
\label{sec:local_clumping}
The global averages compress all small-scale structure into a single number per redshift. To isolate where this structure enters, it is useful to write the recombination rate directly as an integral over phase space. Throughout this section we denote $x \equiv x_\HII$ and use
\begin{equation}
n_{\HII} = x\,\bar{n}_{\rm H}\,\Delta \, ,
\end{equation}
with $\bar n_{\rm H}(z)$ being the mean hydrogen number density corresponding to the density normalization of the saved overdensity bins.

\textit{Volume-weighted PDFs and conditional moments:}
Let $P_V(\Delta,T,x)$ be the fraction of the \emph{total} simulation volume per element of $(\Delta,T,x)$ phase space. Consistent with Eq.~\eqref{eq:selected_gas_moment}, restricting the phase-space domain to $\Delta<100$ gives
\begin{equation}
\int_{\Delta<100} \text{d}\Delta \int \text{d}T \int_{0}^{1} \text{d}x\; P_V(\Delta,T,x) = f_{V,<100} \, ,
\end{equation}
where $f_{V,<100}=\sum_{\Delta_i<100}V_i/\sum_iV_i$ is the selected volume fraction (and is very close to unity here). We do not renormalize $P_V$ by $f_{V,<100}$ when reconstructing the global rate.
We define the $(\Delta,T)$ marginal by integrating out the ionization coordinate,
\begin{equation}
P_V(\Delta,T) \equiv \int_{0}^{1} \text{d}x\; P_V(\Delta,T,x) \, .
\end{equation}
For any function $f(x)$ we then define the conditional moment within a $(\Delta,T)$ bin as
\begin{equation}
\langle f(x)\rangle_{\Delta,T}
\equiv
\frac{1}{P_V(\Delta,T)}
\int_{0}^{1} \text{d}x\; f(x)\,P_V(\Delta,T,x) \, .
\end{equation}
These conditional moments quantify \emph{ionization patchiness} at fixed overdensity and temperature, i.e.~the spread in $x$ among cells that share similar $(\Delta,T)$.

\textit{Recombination rate as a phase-space integral:}
For the phase-space decomposition, we first write the volume-averaged recombination rate density in the hydrogen-only approximation:
\begin{equation}
\Gamma_{\rm rec} \equiv \left\langle \alpha_B(T)\,n_{\HII}^2 \right\rangle_{<100} \, .
\end{equation}
Expressed in full $(\Delta,T,x)$ phase space this becomes
\begin{equation}
\Gamma_{\rm rec}
=
\bar n_{\rm H}^2
\int_{\Delta<100} \text{d}\Delta \int \text{d}T \int_{0}^{1} \text{d}x\;
\alpha_B(T)\,\Delta^2 x^2\,P_V(\Delta,T,x) \, ,
\label{eq:Gamma_rec_3D}
\end{equation}
while integrating out $x$ yields the equivalent $(\Delta,T)$ form
\begin{equation}
\Gamma_{\rm rec}(z)
=
\bar n_{\rm H}^2
\int_{\Delta<100} \text{d}\Delta \int \text{d}T\;
\alpha_B(T)\,\Delta^2 \langle x^2\rangle_{\Delta,T}\,P_V(\Delta,T) \, .
\label{eq:Gamma_rec_2D}
\end{equation}

\textit{Ionization patchiness within phase space $\mathcal{P}_{\rm ion}(\Delta,T,z)$:}
To make explicit how ionization inhomogeneity enters Eq.~\eqref{eq:Gamma_rec_2D}, we rewrite the second ionization moment as
$\langle x^2\rangle_{\Delta,T} = \mathcal{P}_{\rm ion}(\Delta,T,z)\,\langle x\rangle_{\Delta,T}^2$,
where we define the \emph{ionization patchiness factor}
\begin{equation}
\mathcal{P}_{\rm ion}(\Delta,T,z)
\equiv
\frac{\langle x^2\rangle_{\Delta,T}}{\langle x\rangle_{\Delta,T}^2} \, .
\label{eq:Pion_def}
\end{equation}
With this definition, Eq.~\eqref{eq:Gamma_rec_2D} becomes
\begin{equation}
\begin{aligned}
\Gamma_{\rm rec}(z)
&=
\bar n_{\rm H}^2
\int_{\Delta<100} \text{d}\Delta \int \text{d}T\;
\alpha_B(T)\,\Delta^2 \\
&\qquad\times
\mathcal{P}_{\rm ion}(\Delta,T,z)\,
\langle x\rangle_{\Delta,T}^2\,
P_V(\Delta,T,z) \, .
\end{aligned}
\label{eq:Gamma_rec_2D_Pion}
\end{equation}
The motivation for Eq.~\eqref{eq:Pion_def} is direct: if the ionization field were uniform within a given $(\Delta,T)$ bin with nonzero mean ionization, then $x$ would be nearly constant and $\mathcal{P}_{\rm ion}=1$, so $\langle x^2\rangle_{\Delta,T}=\langle x\rangle_{\Delta,T}^2$. Deviations from unity therefore quantify ionization patchiness among gas elements with the same overdensity and temperature. In particular, $\mathcal{P}_{\rm ion}>1$ indicates a broad distribution in $x$, e.g.~partially ionized structures mixed with highly ionized gas at the same $(\Delta,T)$, which boosts recombinations relative to any model that uses only $\langle x\rangle_{\Delta,T}$.

\textit{Phase space clumping factor $\mathcal{C}_{\rm ps}(\Delta,T,z)$:}
We now define a clumping factor that isolates the enhancement of recombinations due purely to small-scale ionization patchiness and temperature, separated from how much volume occupies a given thermodynamic state. We use $\alpha_4 \equiv \alpha_B(10^4\,{\rm K})$ for the Case~B recombination coefficient at $10^4\,{\rm K}$. The guiding requirement is that the hydrogen-only phase-space rate can be written in the homogeneous structural form
\begin{equation}
\begin{aligned}
\Gamma_{\rm rec}(z)
&=
\int \text{d}\Delta \int \text{d}T\;
\alpha_4\,
\mathcal{C}_{\rm ps}(\Delta,T,z) \\
&\qquad\times
\langle n_{\HII}\rangle_{\Delta,T}^2\,
P_V(\Delta,T,z) \, .
\end{aligned}
\label{eq:Gamma_rec_C_loc_def}
\end{equation}
In this definition, $\mathcal{C}_{\rm ps}$ is intended to capture the \emph{boost} to recombinations relative to a uniform-ionization estimate at fixed $(\Delta,T)$, while the factor $P_V(\Delta,T)$ separately accounts for how much gas actually resides in that region of phase space.

Comparing Eq.~\eqref{eq:Gamma_rec_C_loc_def} with the exact expression above immediately yields
\begin{equation}
\mathcal{C}_{\rm ps}(\Delta,T,z) = \frac{\alpha_B(T)}{\alpha_4}\,\mathcal{P}_{\rm ion}(\Delta,T,z) \, .
\label{eq:Cloc_final}
\end{equation}
In this form, $\mathcal{C}_{\rm ps}$ quantifies how recombinations in gas of a given overdensity and temperature are enhanced relative to a baseline model that assumes uniform ionization and a fixed $10^4$\,K recombination coefficient. The two physical ingredients entering $\mathcal{C}_{\rm ps}$ are explicit. The factor $\mathcal{P}_{\rm ion}$ isolates ionization patchiness at fixed $(\Delta,T)$, while the ratio $\alpha_B(T)/\alpha_4$ captures the local thermal reduction or enhancement of recombinations through the temperature dependence of $\alpha_B(T)$.

It is important to distinguish this intrinsic boost at a given point in phase space from its \emph{global impact}. Whether a large value of $\mathcal{C}_{\rm ps}$ contributes significantly to the total recombination budget depends on additional weights: the relative volume $P_V(\Delta,T)$, the density factor $\Delta^2$, and the mean ionized fraction $\langle x\rangle_{\Delta,T}^2$. Collecting these terms, the recombination contribution density in phase space may be written as
\begin{equation}
n_{\rm rec}(\Delta,T,z)
=
\alpha_4\,
\mathcal{C}_{\rm ps}(\Delta,T,z)\,
\bar n_{\rm H}^2\,\Delta^2\,
\langle x\rangle_{\Delta,T}^2\,
P_V(\Delta,T,z)\,,
\label{eq:nrec_phase}
\end{equation}
so that
\begin{equation}
\Gamma_{\rm rec}(z)
=
\int \text{d}\Delta \int \text{d}T\; n_{\rm rec}(\Delta,T,z)\,.
\end{equation}
Defining the recombination probability density as
$P_{\rm rec}(\Delta,T,z)\equiv n_{\rm rec}(\Delta,T,z)/\Gamma_{\rm rec}(z)$, its integral over $(\Delta,T)$ space is unity by construction.

In this decomposition, $\mathcal{P}_{\rm ion}$ isolates the effect of ionization patchiness alone, $\mathcal{C}_{\rm ps}$ captures the enhancement including the temperature dependence of $\alpha_B(T)$, and the remaining factors determine how strongly that enhancement is weighted in the global recombination budget. This separation clarifies why large $\mathcal{C}_{\rm ps}$ does not necessarily imply global importance unless it coincides with sufficiently high density, ionized fraction, and occupied volume.

The numerical value of $\mathcal{C}_{\rm ps}$ depends on the adopted $(\Delta,T)$ discretization because coarser bins mix a wider range of ionization states and can artificially increase the apparent ionization patchiness. For this reason we use $\mathcal{C}_{\rm ps}$ primarily as an interpretive diagnostic of local phase-space structure. The physically relevant quantity for the global recombination budget is the contribution-weighted field $n_{\rm rec}(\Delta,T)$, which combines the intrinsic boost with the density, ionized-fraction, and volume weights that determine the actual recombination rate.

Our definition of the phase-space clumping factor $\mathcal{C}_{\rm ps}$ is complementary to several existing notions of ``clumping'' in the reionization literature, each designed for a different modeling purpose:
\begin{itemize}
    \item \textbf{Phase-space clumping ($\mathcal{C}_{\rm ps}$; this work):}
    $\mathcal{C}_{\rm ps}(\Delta,T,z)$ characterizes the ionization inhomogeneity of gas occupying a given thermodynamic state in phase space. Rather than corresponding to a localized spatial region, it measures the recombination boost arising from ionization patchiness within a $(\Delta,T)$ bin. In this sense, $\mathcal{C}_{\rm ps}$ may be viewed as a form of \emph{phase-space local} clumping, designed for global recombination accounting in phase space.

    \item \textbf{Spatially-local clumping ($\mathcal{C}_{\rm loc}$):}
    Local clumping factors analogous to $\mathcal{C}_{100}$ defined in Eq.~\eqref{C_100 equation} \citep[cf.][]{KaurovGnedin2015,Sobacchi2014,Mao2020,Bianco2021} are constructed by dividing the simulation volume into spatial subregions and measuring the clumping of ionized gas within each region. The resulting $\mathcal{C}_{\rm loc}$ is then correlated with the local density or environment and is primarily intended for local or subgrid radiative-transfer models.

    \item \textbf{Global clumping factors:}
    Many studies employ a single volume-averaged clumping factor (in the same manner as Eqs.~\ref{eq:CHII_def} and \ref{eq:Crec_def}) to correct unresolved recombinations in global ionization-balance equations \citep[e.g.][]{Madau1999,Pawlik2009,RaicevicTheuns2011,JeesonDaniel2014,Shull2012}. These quantities compress all spatial and thermodynamic information into a single number at each redshift.

    \item \textbf{Mean-free-path and absorber-based approaches:}
    Other models emphasize the role of self-shielded absorbers, Lyman-limit systems, and mean-free-path-limiting structures rather than a single clumping factor \citep[e.g.][]{MiraldaEscude2000,FurlanettoOh2005,McQuinn2011,Rahmati2013,Sobacchi2014,DaviesFurlanetto2016}. In these approaches, recombination sinks are characterized through absorber populations and radiative transfer rather than through an effective clumping correction.
\end{itemize}

\subsection{Thermal equilibrium bands and self-shielding}\label{sec:thermal_bands}
We use analytic thermal-equilibrium bands to interpret the recombination-weighted gas distribution in the $(\Delta,T)$ plane. The full derivation from the thermal-energy equation, including the adopted heating and cooling terms and their pair-normalized forms, is given in Appendix~\ref{app:thermal_bands}. The lower photoheated band, $T_{\rm low}(\Delta|z)$, includes radiative, adiabatic, and Compton cooling and is the implicit solution of
\begin{align} \label{eq:Tlow_implicit_full}
  &\alpha_{\rm B}(T)\,\langle E_{\rm heat}\rangle
  =
  \Lambda_{\rm rec}^{\HII}(T)
  + x_{\HI}(T,\Delta)\,\Lambda_{\rm exc}^{\HI}(T)
  \notag\\
  &\quad+
  \left[
  \frac{3k_{\rm B}H(z)}
  {X\mu\,n_{\rm H}(\Delta,z)}
  \right]T
  +
  \left[
  \frac{3}{2}
  \frac{k_{\rm B}K_{\rm C}(z)f_e}
  {X\mu\,n_{\rm H}(\Delta,z)}
  \right](T-T_{\rm CMB}) \, .
\end{align}
Here $n_{\rm H}(\Delta,z)=X\rho_b(z)\Delta/m_\text{H}$, $f_e$ is the electron-to-total-particle ratio, and $K_{\rm C}(z)$ is the Compton coupling coefficient defined in Appendix~\ref{app:thermal_bands}. The upper photoheated band, $T_{\rm high}(\Delta|z)$, is obtained by dropping adiabatic and Compton cooling:
\begin{equation} \label{eq:Thigh_implicit_full}
  \alpha_{\rm B}(T)\,\langle E_{\rm heat}\rangle
  =
  \Lambda_{\rm rec}^{\HII}(T)
  + x_{\HI}(T,\Delta)\,\Lambda_{\rm exc}^{\HI}(T) \, .
\end{equation}
The density dependence enters through $n_{\rm H}(\Delta,z)$ and the equilibrium neutral fraction $x_{\HI}(T,\Delta)\simeq\alpha_{\rm B}(T)n_{\rm H}(\Delta,z)/\Gamma_{\HI}$. We adopt fixed effective photoheating energies $\langle E_{\rm heat}\rangle=3.25\,{\rm eV}$ for \thesanone and \thesantwo and $4.20\,{\rm eV}$ for \lumina. For $T_{\rm low}$ and $T_{\rm high}$ we use $\Gamma_{\HI}=10^{-12}\,{\rm s^{-1}}$. We obtain $T_{\rm SS,high}$ and $T_{\rm SS,low}$ by solving Eq.~\eqref{eq:Thigh_implicit_full} with the reduced local rates $\Gamma_{\HI}=10^{-13}$ and $10^{-16}\,{\rm s^{-1}}$, respectively, which bracket moderate and strong self-shielding \citep[cf.][]{Rahmati2013,KaurovGnedin2015}. Further details and the derivation of all four bands are provided in Appendix~\ref{app:thermal_bands}.

\subsection{Zone-based classification in \texorpdfstring{$(\Delta,T)$}{(Δ,T)}}\label{sec:zones}
Guided by the redshift- and density-dependent thermal equilibrium boundaries
$T_{\rm low}(z,\Delta)$, $T_{\rm high}(z,\Delta)$, and the self-shielding
thresholds $T_{\rm SS,low/high}(z,\Delta)$, we decompose gas in the
$(\Delta,T)$ plane into the following physically motivated zones:
\begin{align}
  \text{PI~IGM:}\quad & T_{\rm low} < T < T_{\rm high}, \; \Delta_{\rm min} < \Delta < 10 \label{eq:zone_pi_igm}\\[6pt]
  \text{LTNE:}\quad & 10^{3}\,{\rm K} < T < T_{\rm low}, \; \Delta < 10 \label{eq:zone_ltne_igm}\\[3pt]
  & 10^{3}\,{\rm K} < T < T_{\rm SS,low}, \; 10 \le \Delta < 100 \label{eq:zone_ltne_cgm}\\[6pt]
  \text{PI~OD:}\quad & T_{\rm SS,high} < T < T_{\rm high}, \; 10 < \Delta < 100 \label{eq:zone_pi_od}\\[6pt]
  \text{SS~OD:}\quad & T_{\rm SS,low} < T < T_{\rm SS,high}, \; 10 < \Delta < 100 \label{eq:zone_ss_od}\\[6pt]
  %
  \text{HTNE:}\quad & T_{\rm high} < T < 10^{6}\,{\rm K}, \; 1 < \Delta < 100 \, . \label{eq:zone_htne}
\end{align}
The photoionized intergalactic medium (PI~IGM) corresponds to low-density gas that has reached photoionization equilibrium, lying between the equilibrium heating and cooling branches in $(\Delta,T)$ space and dominating the diffuse IGM, as described by HG97 and \citet{Schaye1999}; $\Delta_{\rm min}=0.1$ is the minimum overdensity in our analysis (the plot minimum). The low-temperature non-equilibrium (LTNE) zone captures gas below the photoionization-equilibrium temperature. This category selects thermally evolving gas whose thermal state can reflect adiabatic expansion and cooling, pressure-smoothing response, shocks and other local hydrodynamic processes, and its prior ionization and heating history
\citep{TracCenLoeb2008,McQuinnUptonSanderbeck2016,McQuinn2016,Daloisio2018}. We refer to the two $10<\Delta<100$ zones as photoionized overdense gas (PI~OD) and self-shielded overdense gas (SS~OD), since they include dense filaments and non-star-forming minihaloes as well as halo-associated gas. PI~OD remains optically thin to the ionizing background and occupies temperatures between the cooler self-shielded equilibrium ($T_{\rm SS,high}$) and the optically thin heating-limited branch ($T_{\rm high}$). SS~OD traces dense gas where recombination and cooling dominate over photoheating, leading to partial or full self-shielding and lower equilibrium temperatures \citep{Rahmati2013,Sobacchi2014}. Finally, the high-temperature non-equilibrium (HTNE) zone represents gas that has been heated to high temperatures by gravitational collapse, accretion shocks, or feedback processes, independent of photoionization equilibrium, and contributes to recombinations primarily at high overdensities.

For each zone $Z$, we compute the fractional contribution to the instantaneous hydrogen recombination rate as $f_\text{rec}(Z) = \dot{N}_\text{rec}(Z) / \dot{N}_\text{rec}^{\rm tot}$, where the numerator is the rate from zone $Z$ and the denominator is the total rate from all gas, in both cases restricting to $\Delta < 100$.

\section{Results}\label{sec:results}

\subsection{Global clumping prescriptions and photon budget}
\label{sec:clumping_evolution}
\begin{figure}
    \centering
    \safeincludegraphics[width=\columnwidth]{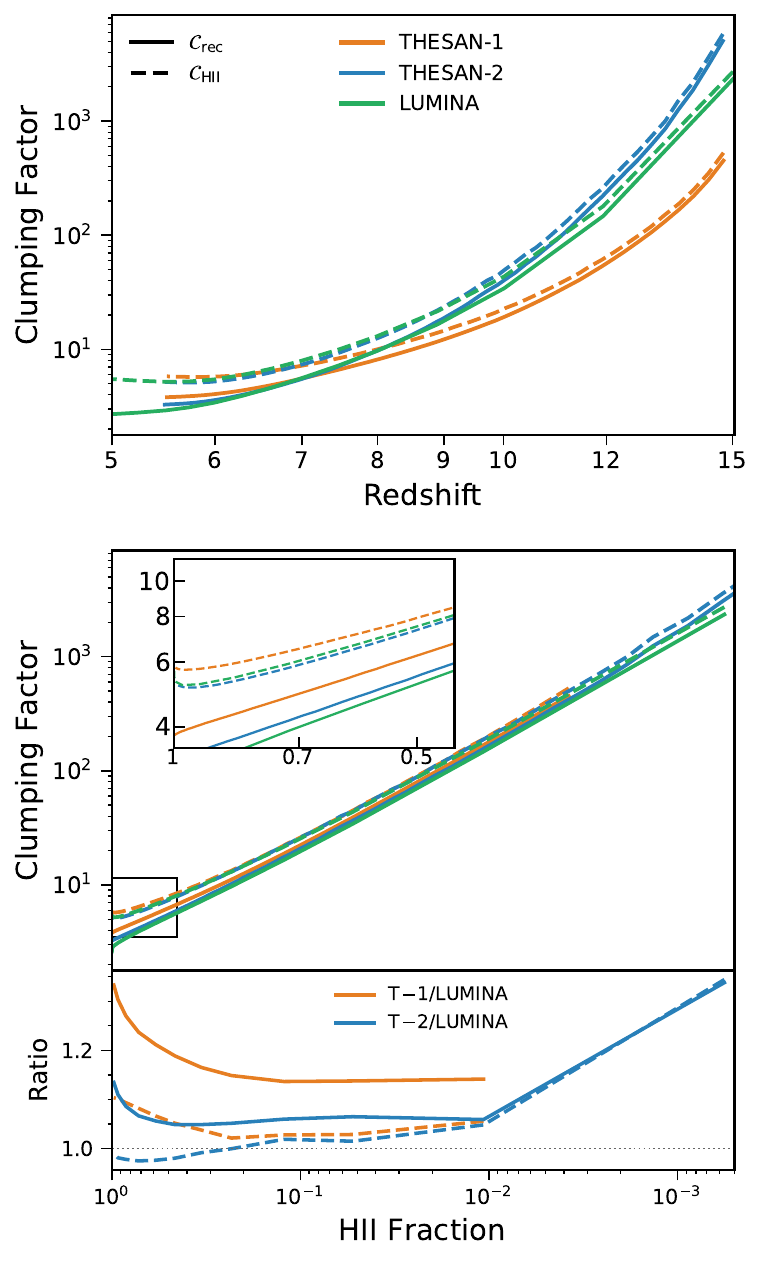}
    \caption{\textit{Top panel:} Evolution of global clumping factors whose numerators include gas with overdensity $\Delta<100$ and whose denominator densities are averaged over all gas in \thesanone, \thesantwo, and \lumina, plotted as a function of logarithmic redshift. The redshift axis follows the convention used throughout the paper, with high redshift on the right. Solid curves show the recombination-weighted clumping factor $\mathcal{C}_{\rm rec}$, while dashed curves show the density-only clumping factor $\mathcal{C}_{\rm \HII}$. \textit{Bottom panel:} The same clumping curves plotted as a function of the global \HII fraction, using the simulation-specific reionization histories to map $(z \rightarrow x_{\rm \HII})$. Expressed in this form, the three simulations agree at the $\sim 10$--$20\%$ level over most of the reionization history. The inset highlights the late stages of reionization ($x_{\rm \HII} \gtrsim 0.5$). The ratio region within the bottom panel shows the clumping factors in \thesanone and \thesantwo relative to \lumina; both $\mathcal{C}_{\rm rec}$ (solid) and $\mathcal{C}_{\rm \HII}$ (dashed) remain within $\sim 10$--$20\%$ of the \lumina values over most of the reionization history. Appendix~\ref{app:fits_cxhii} gives compact power-law fits to these relations.}
    \label{fig:clumping-compare}
\end{figure}

Figure~\ref{fig:clumping-compare} compares the ionized-hydrogen density-only clumping factor $\mathcal{C}_\text{\HII}$ with the recombination-weighted clumping factor $\mathcal{C}_\text{rec}$ across the three simulations. In both quantities the numerator isolates the $\Delta<100$ sink contribution, whereas the denominator uses the corresponding all-gas mean densities; the plotted factors can therefore multiply the global density terms in a photon-counting model without an additional volume renormalization. The most robust trend is the consistent offset between the two, with $\mathcal{C}_\text{rec} < \mathcal{C}_\text{\HII}$ in all three runs, and the separation growing toward later times. This is a result of photoheated ionized gas recombining less efficiently than a medium with a fixed temperature of $10^4\,\mathrm{K}$. When the same factors are plotted against the global ionized fraction, the three simulations follow an approximately common relation at the $\sim10$--$20\%$ level. We therefore regard $x_{\HII}$ as a more physically meaningful variable for recombination efficiency than redshift alone. The residual scatter shows that the relation is not strictly universal. Appendix~\ref{app:fits_cxhii} provides power-law fits to these $\mathcal{C}(x_{\HII})$ trends that may be useful for semi-analytic models.

At fixed redshift during the early and intermediate stages of reionization ($z\gtrsim7$), the simulations are offset from one another in a consistent ordering: \thesanone yields the smallest clumping, \thesantwo the largest, and \lumina lies between them; toward the end of reionization this ordering inverts, with \thesanone exhibiting the largest clumping (see the ratio panel of Figure~\ref{fig:clumping-compare}). The particularly close agreement between the $\mathcal{C}_{\rm \HII}$ curves of \thesantwo and \lumina is consistent with their similar mass resolution, indicating that the much larger \lumina volume introduces only modest changes to this statistic. The remaining differences plausibly reflect the updated radiation treatment and other physics in \lumina, together with its different reionization history.

\begin{figure}
    \centering
    \safeincludegraphics[width=\columnwidth]{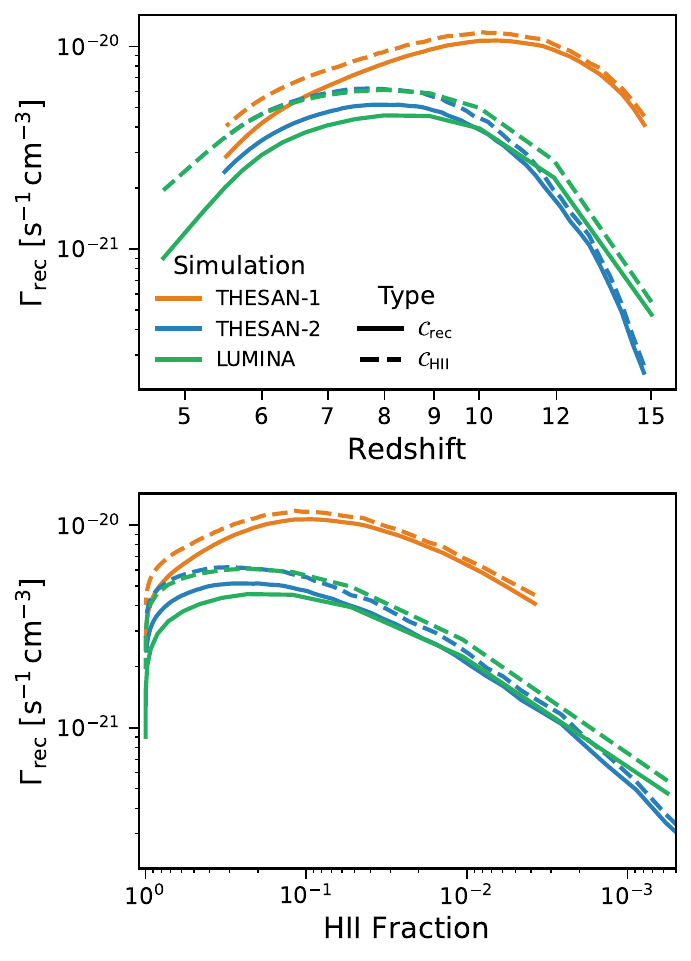}
    \caption{\textit{Top:} Volume-averaged recombination rate density in the three simulations. Solid (dashed) curves correspond to the $\mathcal{C}_\text{rec}$ ($\mathcal{C}_\text{\HII}$) prescriptions. All models show a broad maximum in recombination rate around $z\sim8$--$10$, followed by a decline toward $z\sim5$. In \lumina, the $\mathcal{C}_\text{\HII}$-based rate exceeds the $\mathcal{C}_\text{rec}$ rate by factors of $1.34$ at $z\simeq8$ and $2.02$ at $z\simeq5$. Among the simulations, \thesanone reaches the highest peak rate, followed by \thesantwo, with \lumina lowest. \textit{Bottom:} The same recombination rates plotted against \HII fraction. Unlike the clumping factors, the recombination-rate curves do not collapse across simulations. This indicates that matching $x_{\HII}$ and $\mathcal{C}_{\rm rec}$ does not fix the density term in $\Gamma_{\rm rec}\propto \mathcal{C}_{\rm rec}\langle n_{\HII}\rangle^2$. The higher \thesanone values are consistent with denser ionized gas being sampled, e.g., an earlier reionization history combined with more resolved radiation sources and sinks.}
    \label{fig:rec-rate}
\end{figure}

Figure~\ref{fig:rec-rate} presents the corresponding recombination rate densities computed using both $\mathcal{C}_\text{\HII}$ and $\mathcal{C}_\text{rec}$. Plotted against redshift (top panel), the rates in all simulations rise steeply between $z\approx15$ and $z\approx8$--$10$, reach a broad maximum, and then decline toward $z\approx5$. As with the clumping factors, the $\mathcal{C}_\text{rec}$-based rates closely track the $\mathcal{C}_\text{\HII}$ curves but remain consistently lower. For \lumina, the density-only prescription overpredicts the rate by factors of $1.34$ at $z\simeq8$ and $2.02$ at $z\simeq5$. However, unlike the clumping factors, the rates also retain the absolute density scale entering the recombination integral: \thesanone produces the highest recombination rate, followed by \thesantwo, with \lumina lowest.

When recombination rates are plotted against ionized fraction (bottom panel), the curves no longer collapse across simulations. This is expected, since $\Gamma_{\rm rec}\propto \mathcal{C}_{\rm rec}\,\langle n_{\HII}\rangle^2$, implying that matching the clumping factor at fixed $x_{\rm \HII}$ does not by itself enforce the same characteristic density of ionized gas. The higher $\Gamma_{\rm rec}$ values in \thesanone are therefore consistent with ionization sampling denser gas than in \lumina at the same global ionized fraction, but we do not directly isolate the origin of that difference.

\begin{figure}
    \centering
    \safeincludegraphics[width=\columnwidth]{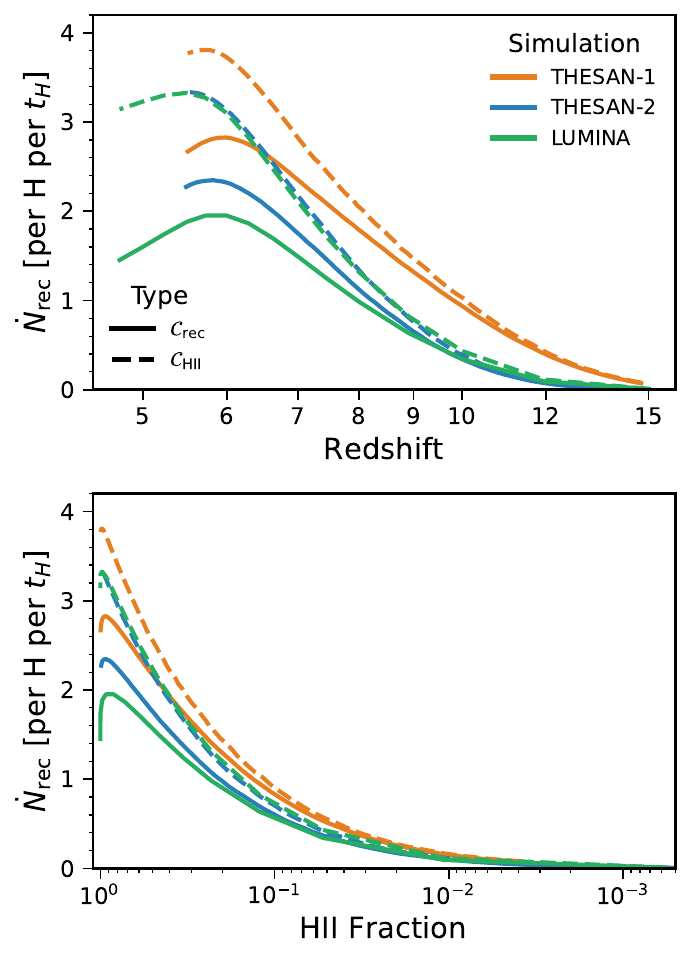}
    \caption{\textit{Top:} Number of recombinations per hydrogen atom per Hubble time, $\dot{N}_\text{rec}$, as a function of redshift for the three simulations. Solid and dashed curves correspond to the $\mathcal{C}_\text{rec}$ and $\mathcal{C}_\text{\HII}$ prescriptions, respectively. Values grow from $\ll1$ at high redshift to several recombinations per atom by $z\simeq6$. For \lumina, the density-only value exceeds the temperature-weighted value by factors of $1.34$ at $z\simeq8$ and $2.02$ at $z\simeq5$. \textit{Bottom:} The same per-Hubble-time recombination counts plotted as a function of \HII fraction. As with the instantaneous rates, the curves do not overlap across simulations. The ionization-fraction parametrization removes much of the difference in clumping growth but does not eliminate differences tied to the density of ionized gas.}
    \label{fig:Nrec-compare}
\end{figure}

Figure~\ref{fig:Nrec-compare} shows the number of recombinations per hydrogen atom per Hubble time, $\dot{N}_\text{rec}(z)$. The overall shape of the curves parallels that of the instantaneous rates: $\dot{N}_\text{rec}$ increases from negligible values at $z\gtrsim15$ to of order unity by $z\sim8$, and reaches several recombinations per atom by $z\sim6$. As before, the $\mathcal{C}_\text{\HII}$-based curves lie above the $\mathcal{C}_\text{rec}$ curves for all simulations; for \lumina the ratios are $1.34$ at $z\simeq8$ and $2.02$ at $z\simeq5$. At fixed redshift \thesanone consistently yields the largest per-Hubble-time recombination count, followed by \thesantwo and then \lumina. When expressed as a function of ionized fraction, the results from the different simulations do not converge onto a single relation. This is consistent with the contribution of the underlying density field: even at identical global ionization levels, the factor $\propto n_\text{H}^2$ weights denser ionized gas disproportionately.

\begin{figure}
    \centering
    \safeincludegraphics[width=\columnwidth]{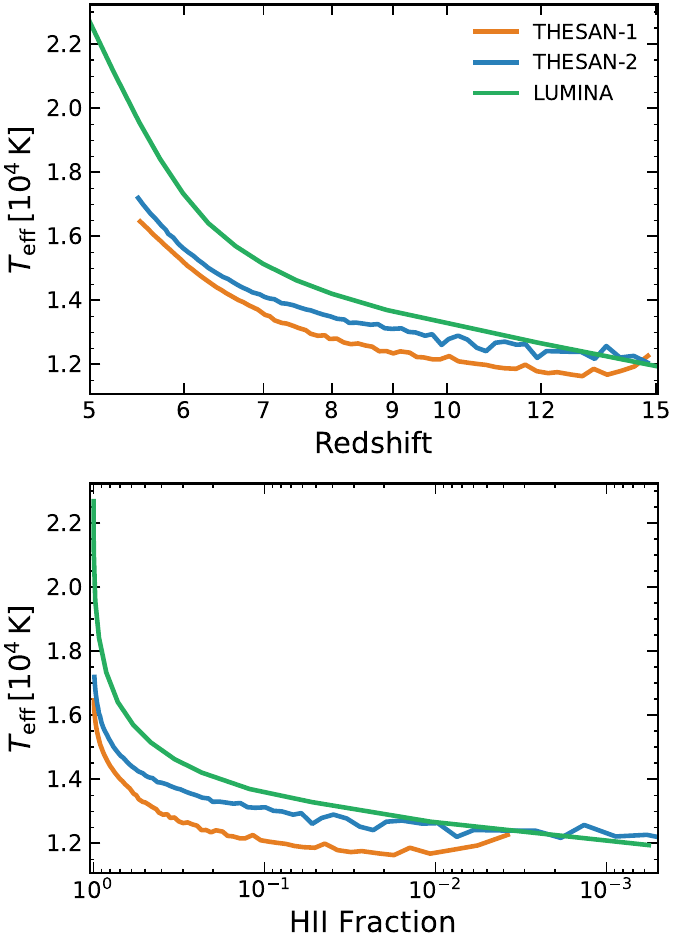}
    \caption{\textit{Top:} Effective recombination temperature $T_\text{eff}$ implied by the full recombination rate, defined via Eq.~\eqref{eq:Teff_ratio_final}, as a function of redshift. \textit{Bottom:} The same temperatures plotted against the global \HII fraction using each simulation's reionization history. All simulations show a smooth evolution from $(1.2$--$1.3)\times 10^4\,\mathrm{K}$ at high redshift to $(1.7$--$2.3)\times 10^4\,\mathrm{K}$ near $z\simeq5$. \thesanone yields the lowest $T_\text{eff}$, \thesantwo slightly higher values, and \lumina the highest over most of the redshift range.}
    \label{fig:Teff}
\end{figure}

The effective temperatures implied by these global rates are presented in Figure~\ref{fig:Teff}. For each simulation and redshift we solve Eq.~\eqref{eq:Teff_ratio_final} for $T_\text{eff}(z)$, i.e.~the single temperature that, combined with the electron--ion density factor $\mathcal{C}_{e\HII}(z)$, reproduces the exact $\Gamma_\text{rec}(z)$ as if $\mathcal{C}_\text{rec}(z)$ had been used all along. The top panel shows the redshift evolution, while the bottom panel compares the same values at fixed global ionized fraction. All three simulations show a smooth, broadly monotonic evolution: $T_\text{eff}$ decreases from $\sim(1.7$--$2.3)\times 10^4\,\mathrm{K}$ at $z\simeq5$ down to $\sim(1.2$--$1.3)\times10^4\,\mathrm{K}$ by $z\simeq14$. At most redshifts \thesanone has the lowest $T_\text{eff}$, \thesantwo is slightly hotter, and \lumina is hottest, with the curves converging at $z\gtrsim13$. This ordering is physically consistent with the additional hard radiation in \lumina from accreting black holes, high-mass X-ray binaries, and hot ISM gas, components absent from \thesan that provide additional heating. Throughout the interval $T_\text{eff}$ remains above $10^4\,\mathrm{K}$ in all three runs.

\begin{figure}
    \centering
    \safeincludegraphics[width=\columnwidth]{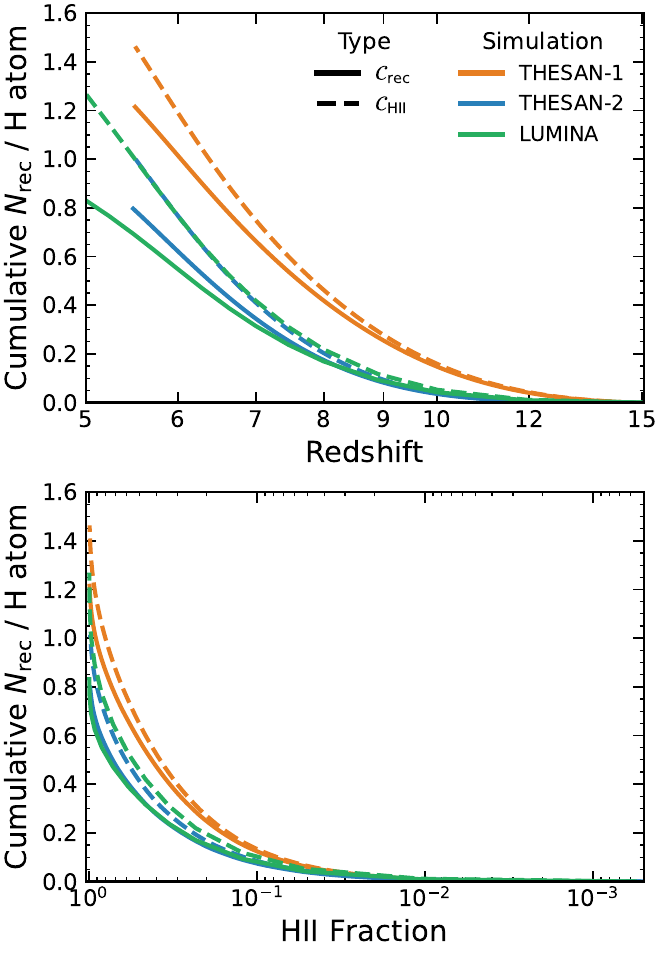}
    \caption{\textit{Top:} Cumulative number of recombinations per hydrogen atom from $z_{\rm max}\simeq 15$ down to each redshift $z$. \textit{Bottom:} The same cumulative counts plotted against global \HII fraction. Solid curves use $\mathcal{C}_\text{rec}$ and dashed curves use $\mathcal{C}_\text{\HII}$. At all stages the $\mathcal{C}_\text{\HII}$-based integrals exceed their $\mathcal{C}_\text{rec}$ counterparts, and the difference increases with time. In \lumina, by $z\simeq5$ the corresponding values are $1.269$ and $0.831$ recombinations per hydrogen atom, a ratio of $1.53$.}
    \label{fig:photon_budget}
\end{figure}

Finally, Figure~\ref{fig:photon_budget} shows the cumulative number of recombinations per hydrogen atom between an initial redshift $z_{\rm max}\simeq 15$ and a given redshift $z$, which directly sets the ionizing photon budget required to sustain reionization. Contributions from $z>15$ are negligible in these simulations, so this starting point captures essentially all of the relevant recombination history. We plot separate curves for the $\mathcal{C}_\text{\HII}$ and $\mathcal{C}_\text{rec}$ prescriptions in each simulation, highlighting the impact of temperature-dependent recombination physics on the total photon consumption. The top panel shows the accumulation with redshift, and the bottom panel shows the same cumulative values against global ionized fraction. All curves start from zero at $z=z_{\rm max}$ and grow steadily toward lower redshift, reflecting the cumulative effect of recombinations as ionized regions expand and large-scale structure continues to undergo gravitational collapse. In all three runs, the $\mathcal{C}_\text{\HII}$-based curves lie systematically above their $\mathcal{C}_\text{rec}$ counterparts across the full redshift range, with the separation increasing toward lower redshift. For \lumina, by $z\simeq5$ the cumulative values are $1.269$ and $0.831$ recombinations per hydrogen atom, respectively, so the density-only prescription overpredicts the cumulative budget by a factor of $1.53$. Comparing simulations at fixed redshift, \thesanone yields the largest cumulative recombination budget, followed by \thesantwo, then \lumina.

\subsection{Recombination and volume statistics}
\begin{figure}
    \centering
    \safeincludegraphics[width=\columnwidth]{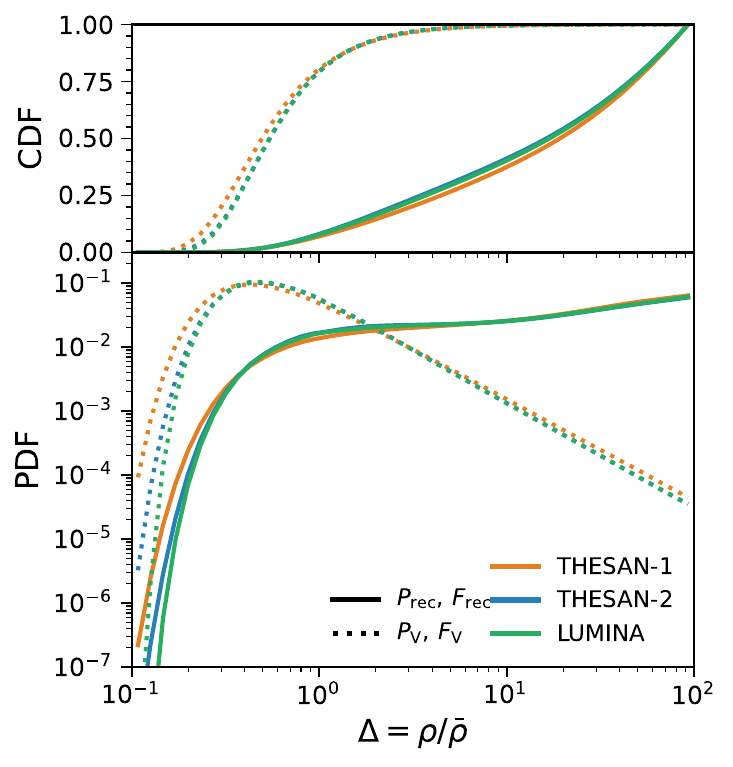}
    \caption{Integrated-over-threshold (top) and differential (bottom) overdensity distributions at $z\simeq 5.99$ for gas with $\Delta<100$. At these snapshots, the global ionized fractions are $x_{\HII}=0.927$, $0.880$, and $0.843$ for \thesanone, \thesantwo, and \lumina, respectively. Solid curves show the recombination-weighted quantities $P_\text{rec}(\Delta)$ and $F_\text{rec}(\Delta)$, while dotted curves show the corresponding volume-weighted statistics $P_V(\Delta)$ and $F_V(\Delta)$. Here $P_\text{rec}$ and $P_V$ are normalized distributions over overdensity bins, and $F_\text{rec}$ and $F_V$ are their cumulative fractions as the overdensity threshold is increased. Across all three simulations, recombinations are overwhelmingly concentrated in overdense regions: more than $80\%$ occur at $\Delta>2$, which occupies less than $10\%$ of the IGM volume, and roughly $60\%$ originate in the high-density tail $\Delta>10$, which occupies under $1\%$ of the volume.}
    \label{fig:pdf-z6}
\end{figure}

At $z\simeq 6$ we examine how recombinations are distributed across the IGM by directly comparing the recombination-weighted and volume-weighted overdensity statistics. The nearest analyzed snapshots are at $z\simeq5.99$, where the global ionized fractions are $x_{\HII}=0.927$, $0.880$, and $0.843$ for \thesanone, \thesantwo, and \lumina, respectively. Figure~\ref{fig:pdf-z6} shows the differential PDFs, $P_\text{rec}(\Delta)$ and $P_V(\Delta)$, and their cumulative counterparts, $F_\text{rec}(\Delta)$ and $F_V(\Delta)$. All three simulations exhibit the same qualitative structure despite their different global ionization states. The volume PDF peaks near $\Delta\sim 0.5$--$1$ and drops sharply toward high densities: only $\sim 1\%$ of the total volume lies at $\Delta>10$. In contrast, the recombination PDF rises steeply to $\Delta\sim 2$ and then increases only mildly toward the $\Delta<100$ boundary.

This separation produces a pronounced divergence between the cumulative curves: although only $\lesssim 10\%$ of the IGM volume resides above $\Delta\simeq 2$, more than $80\%$ of all recombinations originate in this overdense subset. Likewise, roughly $60\%$ of recombinations are produced in the extreme tail $10\le\Delta<100$, which occupies far less than $1\%$ of the total volume. Across the three simulations, the amplitudes differ slightly, but the underlying physical picture is identical: recombinations during reionization are overwhelmingly dominated by a very small and spatially clustered fraction of the cosmic gas.

\begin{figure}
    \centering
    \safeincludegraphics[width=\columnwidth]{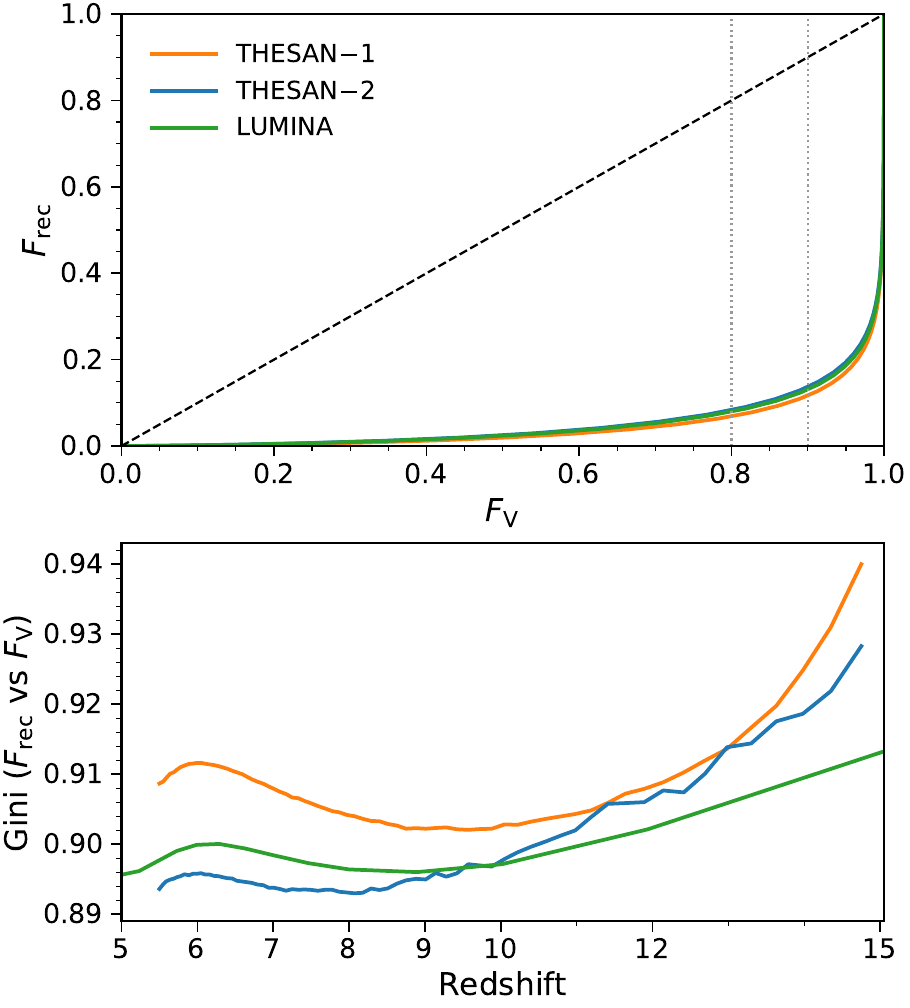}
    \caption[Lorenz curves and Gini coefficients]{Lorenz curves describing the spatial concentration of hydrogen recombinations within gas of overdensity $\Delta<100$ in the \thesantwo, \thesanone, and \lumina simulations. \textit{Top panel:} Comparison of the cumulative recombination fraction $F_\text{rec}$ as a function of cumulative volume fraction $F_V$ at $z\simeq 6$. The dashed line indicates perfect spatial uniformity, while the dotted vertical lines at $F_V=0.8$ and $F_V=0.9$ highlight the strong concentration of recombinations within the densest regions. Across all three simulations, the upper $\sim10\%$ of the volume already accounts for $\gtrsim80\%$ of the total recombination rate. \textit{Bottom panel:} Redshift evolution of the corresponding Gini coefficient, quantifying the degree of recombination clustering. Recombinations are highly concentrated at all redshifts, with only modest differences between the three simulations and a weak evolution from $z\sim15$ to $z\sim5$.}
    \label{fig:lorenz-allz}
\end{figure}

To quantify the degree of spatial concentration of recombinations and its evolution, Figure~\ref{fig:lorenz-allz} presents a complementary two-panel comparison. The top panel shows the Lorenz curves $F_\text{rec}(F_V)$ at $z\simeq 6$ for the \thesanone, \thesantwo, and \lumina simulations. In all three cases, the curves lie far below the one-to-one line, suggesting that recombinations are strongly localized within a small fraction of the total volume. At this redshift, the upper $\sim10\%$ of the volume already accounts for $\sim80$–$90\%$ of the total recombination rate, with only modest differences between the three simulations.

The redshift evolution of the corresponding Gini coefficient provides a single scalar measure of recombination inequality. Across all three simulations, the Gini coefficient remains high ($G \gtrsim 0.9$) over the entire redshift range $5 \lesssim z \lesssim 15$, indicating that recombinations are persistently dominated by dense subregions throughout reionization. In fact, the coefficient would be even more extreme if the restriction to gas with $\Delta < 100$ were removed. The weak redshift evolution reflects a gradual reduction in concentration as ionized regions expand to fill more voids, while the close agreement between simulations highlights the robustness of this behavior to resolution and radiation-field differences. Together, these results suggest that hydrogen recombinations act as highly localized photon sinks at all stages of reionization, a simple but important insight for accurately modeling the ionizing photon budget.

These findings reinforce the well-established picture from analytic and simulation work that dense cosmic structures such as filaments, halo outskirts, and self-shielding regions act as the dominant sinks of ionizing photons during reionization \citep[e.g.][]{MiraldaEscude2000, Finlator2012, Sobacchi2014, DaviesFurlanetto2016}. Our explicit $P_\text{rec}/P_V$ comparisons and Lorenz-curve diagnostics provide quantitative, redshift-resolved measurements of this inhomogeneity.

\begin{figure*}
    \centering
    \safeincludegraphics[width=\textwidth]{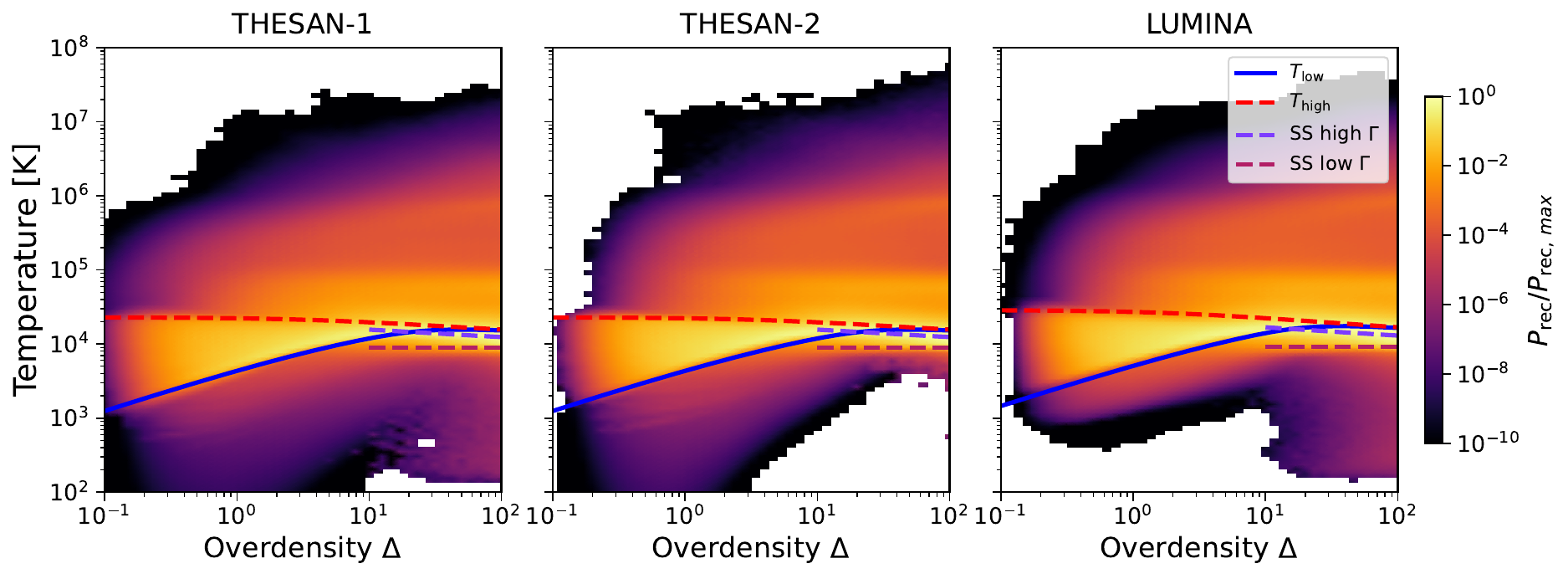}
    \caption[Recombination phase-space distributions]{Recombination probability distribution $P_\text{rec}(\Delta, T)$ in \thesanone, \thesantwo, and \lumina at $z\simeq6$, shown as a function of overdensity and temperature. Each panel displays the corresponding analytic thermal structure: the lower equilibrium branch $\Tlow(\Delta)$, the upper heating-limited branch $\Thigh(\Delta)$, and the self-shielded limits $T_{\rm SS}(\Delta)$ evaluated for high and low photoionization rates. The color scale shows the raw fractional contribution of each $(\Delta,T)$ bin to the global recombination budget, normalized to the maximum; the same color scale is used across the three panels. Across all simulations, recombinations are concentrated in a narrow high-temperature band near $\sim10^4\,\mathrm{K}$, with the width and location of the thermally allowed region varying mildly among runs. Differences between simulations are consistent with differing reionization histories and radiation fields.}
    \label{fig:heatmaps}
\end{figure*}

\begin{figure*}
    \centering
    \includegraphics[width=\textwidth]{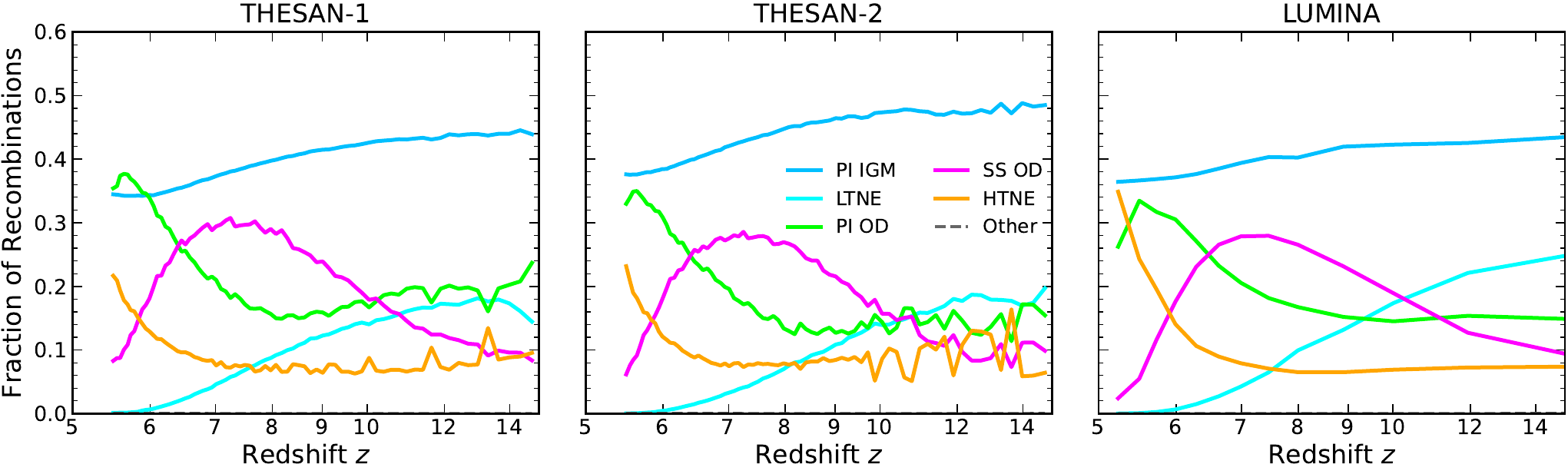}
   \caption[Redshift evolution of recombination-zone contributions]{
Redshift evolution of the fractional contribution to hydrogen recombinations for gas with overdensity $\Delta < 100$ in \thesanone, \thesantwo, and \lumina. The gas is decomposed into photoionized IGM (PI~IGM), low-temperature non-equilibrium gas (LTNE), photoionized overdense gas (PI~OD), self-shielded overdense gas (SS~OD), high-temperature non-equilibrium gas (HTNE), and a small residual ``other'' component. The LTNE category is extended to include overdensities $10 \le \Delta \le 100$ at temperatures below the self-shielding threshold. The PI~IGM contribution is largest at early times and declines as denser phases become more important. SS~OD develops a pronounced intermediate-redshift peak. PI~OD grows toward lower redshift and becomes comparable to, or in some cases larger than, PI~IGM. Hotter non-equilibrium gas also becomes increasingly important at low redshift, particularly in \lumina.
}
\label{fig:zone_evolution}
\end{figure*}

\subsection{Resolving sinks in \texorpdfstring{$(\D,T)$}{(Δ,T)}: thermal bands and self-shielding}
Figure~\ref{fig:heatmaps} shows the recombination probability distribution $P_\text{rec}(\Delta,T)$ at $z\simeq 6$ for all three simulations, together with the analytic thermal bands $\Tlow(\Delta)$, $\Thigh(\Delta)$, and the self-shielded temperatures $T_{\rm SS}(\Delta)$ defined in Section~\ref{sec:thermal_bands} and derived in Appendix~\ref{app:thermal_bands}. The characteristic widening of the thermally allowed region toward low densities reflects the growing importance of adiabatic cooling (with Compton cooling providing a smaller, secondary contribution), which separates the lower and upper equilibrium branches. At high overdensities the gas transitions into the self-shielded regime, where the reduced local photoionization rate lowers the equilibrium temperature and enhances the recombination efficiency, in agreement with previous studies \citep{Rahmati2013, Sobacchi2014, KaurovGnedin2015}. Although the detailed density distributions differ among \thesanone, \thesantwo, and \lumina, the underlying thermal structure of recombination sinks is similar. Recombination-weighted gas occupies a narrow locus near $\sim10^4\,\mathrm{K}$, and the lower thermal boundary has a similar logarithmic slope in all three simulations, ${\rm d}\ln T_{\rm low}/{\rm d}\ln\Delta \approx 0.5$. Thus, the main thermal balance setting the recombination ridge appears to be broadly shared, while the remaining differences are consistent with variations in reionization history and radiation-field topology, although we do not separate those effects.

The two \thesan\ panels look visually similar although they have different recombination budgets. Each phase-space distribution is normalized to its own integral and displayed relative to its own maximum. At $z\simeq6$, gas with $\Delta>10$ supplies $63.8\%$ of recombinations in \thesanone\ and $60.0\%$ in \thesantwo, while the direct Case~B volume-averaged rate is $21\%$ higher in \thesanone. Higher resolution therefore produces a modest shift toward the densest selected gas and a larger absolute normalization, while the common atomic thermal balance preserves the similar $T\sim10^4\,{\rm K}$ ridge.

\begin{figure}
    \centering
    \safeincludegraphics[width=\columnwidth]{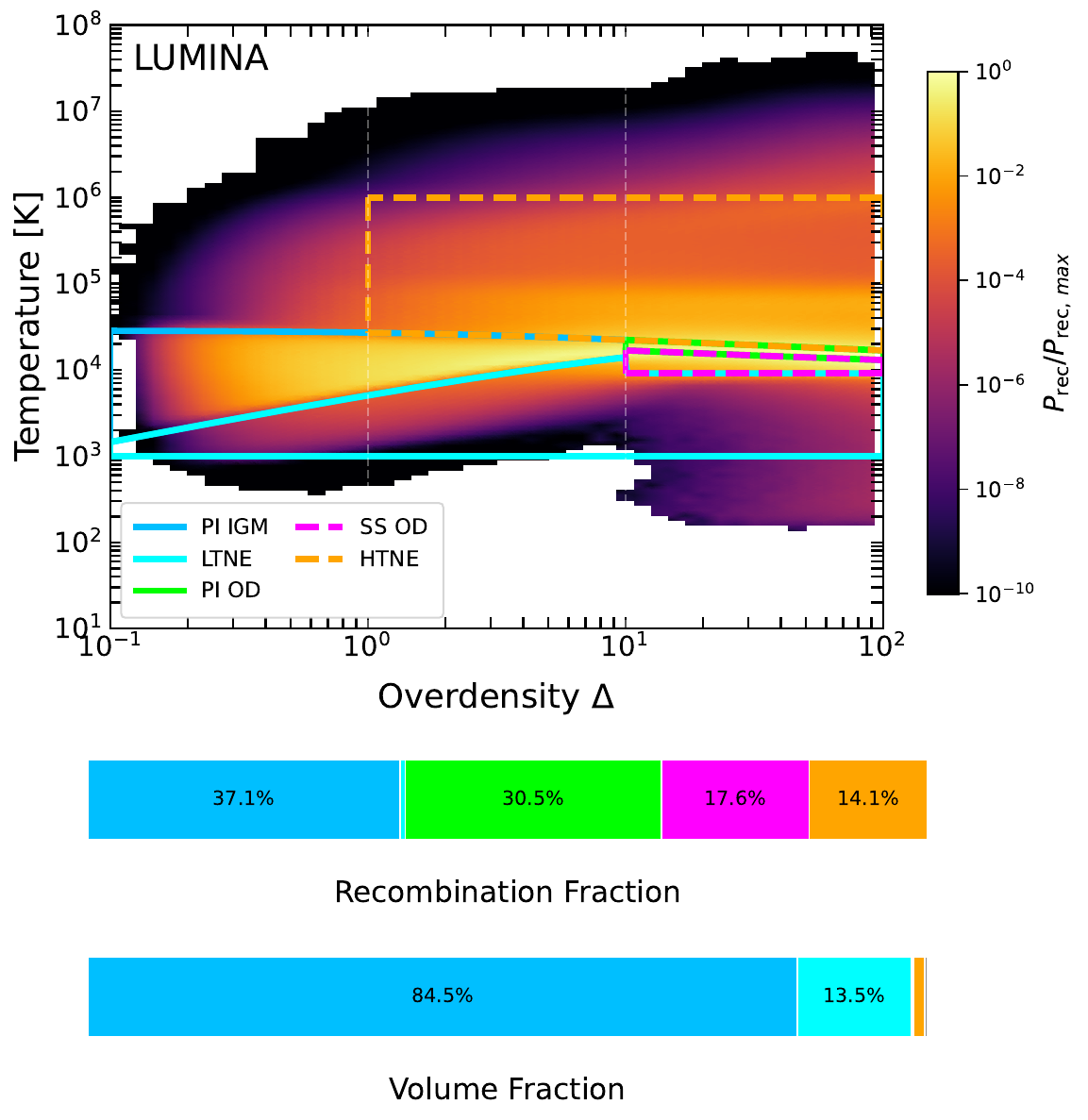}
    \caption[Phase-space decomposition of hydrogen recombinations in LUMINA at z = 6.]{
    Phase-space decomposition of hydrogen recombinations in the \lumina simulation at $z\simeq6$. The main diagram shows the normalized recombination probability $P_{\rm rec}/P_{{\rm rec},\,\max}$ in the overdensity--temperature $(\Delta,T)$ plane for gas with $\Delta<100$, together with analytic thermal boundaries defining the photoionized IGM (PI~IGM), low-temperature non-equilibrium gas (LTNE), photoionized overdense gas (PI~OD), self-shielded overdense gas (SS~OD), and high-temperature non-equilibrium gas (HTNE). The stacked bars beneath the diagram show each zone's fraction of the total recombination rate and occupied volume using identical colors. PI~IGM is the largest single recombination contributor ($37.1\%$) and dominates the volume ($84.5\%$). In contrast, PI~OD and SS~OD together occupy only $0.34\%$ of the volume but contribute $48.1\%$ of recombinations. HTNE contributes $14.1\%$ of recombinations from $1.29\%$ of the volume, whereas LTNE occupies $13.5\%$ of the volume but contributes only $0.66\%$ of recombinations.
    }
    \label{fig:zones}
\end{figure}

\subsection{Zone fractions at \texorpdfstring{$z \simeq 6$}{z = 6} and their redshift evolution}
\label{sec:zone-fractions}
Figure~\ref{fig:zones} presents a detailed phase-space decomposition of hydrogen recombinations in the \lumina simulation at $z\simeq6$, restricted to gas with overdensity $\Delta < 100$. The main diagram shows the recombination probability $P_\text{rec}$ in the $(\Delta,T)$ plane, together with the analytic thermal boundaries that define the different gas phases used throughout this work. At this epoch, recombinations are concentrated in a narrow temperature band around $T \sim 10^{4}\,\mathrm{K}$, reflecting the balance between photoheating and atomic cooling in ionized gas. The stacked bars expose the strong contrast between recombination and volume weighting. PI~IGM is the largest single recombination contributor at $37.1\%$ and fills $84.5\%$ of the selected volume, while LTNE fills another $13.5\%$ but produces only $0.66\%$ of recombinations. Conversely, the dense PI~OD and SS~OD zones occupy only $0.34\%$ of the volume together yet produce $48.1\%$ of recombinations; HTNE similarly contributes $14.1\%$ of recombinations from only $1.29\%$ of the volume. The zone decomposition relates to the PDF and CDF comparison in Figure~\ref{fig:pdf-z6}, in that a very small dense-gas volume carries a disproportionate share of the recombination budget.

The redshift evolution of these zone fractions is shown in Figure~\ref{fig:zone_evolution} for the \thesanone, \thesantwo, and \lumina simulations. Across all three runs, the contribution from PI~IGM decreases toward lower redshift, reflecting the progressive migration of recombination activity from the diffuse IGM into denser environments as ionized regions grow and overlap. At the same time, PI~OD grows and becomes comparable to, or in some runs larger than, PI~IGM at late times; in \lumina at $z\simeq6$, however, PI~IGM remains the largest single zone. SS~OD exhibits a pronounced intermediate-redshift peak. HTNE gas remains subdominant at early times but rises rapidly toward lower redshifts, particularly in \lumina, although we do not isolate whether this reflects halo-mass demographics, feedback energetics, or their coupling to the radiation field.

In Figure~\ref{fig:zone_evolution}, the LTNE category is defined broadly to include thermally evolving gas colder than the photoionization-equilibrium temperature; its temperature can retain the combined influence of adiabatic expansion and cooling, pressure smoothing, shocks, and prior ionization and heating. This phase is most important at early times, contributing more than $20\%$ of the total recombinations at $z\simeq14$ in \lumina, but declines as reionization progresses (negligible by $z\lesssim7$) and an increasing fraction of the IGM is heated toward the equilibrium thermal bands. The prominence of LTNE gas at high redshift therefore shows that cool, thermally evolving material can significantly influence the global recombination budget during the early stages of reionization. Overall, the phase fractions indicate a gradual shift of recombination activity from diffuse LTNE and PI~IGM gas at early times toward denser and more structured environments as reionization proceeds.

\begin{figure}
    \centering
    \safeincludegraphics[width=\columnwidth]{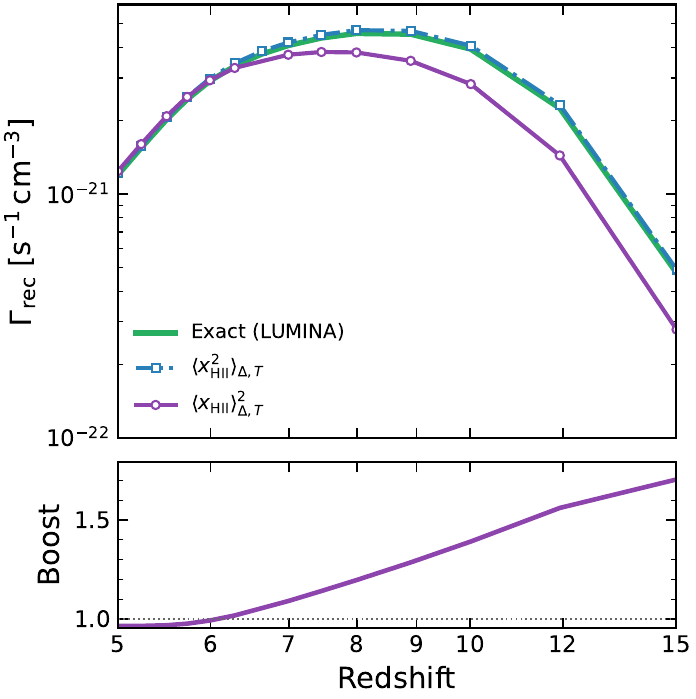}
    \caption[Phase-space estimate of the volume-averaged recombination rate density in LUMINA.]{
    \textit{Top:} Volume-averaged hydrogen recombination rate density in \lumina as a function of redshift. The green curve shows the direct Case~B simulation rate. The purple curve shows the phase-space estimate obtained by replacing the second ionization moment with the squared first moment, $\langle x^2\rangle_{\Delta,T}\rightarrow \langle x\rangle_{\Delta,T}^2$, including the helium-electron correction $\chi_e=1.08$; it underpredicts recombinations when the ionization field is patchy within $(\Delta,T)$ bins. The blue curve uses the binned estimate of the second moment, or equivalently includes $\mathcal{P}_{\rm ion}=\langle x^2\rangle_{\Delta,T}/\langle x\rangle_{\Delta,T}^2$, and closely tracks the exact global rate. \textit{Bottom:} Ratio of the exact rate to the $\langle x\rangle_{\Delta,T}^2$ approximation, directly quantifying the patchiness boost. The enhancement is $1.70$ at early times and approaches unity near overlap.}
    \label{fig:Gamma_rec_Pion_boost}
\end{figure}

\subsection{Phase space clumping and ionization patchiness}
\label{sec:results_local_clumping}

The phase-space decomposition introduced in Section~\ref{sec:local_clumping} provides a direct way to connect ionization patchiness at fixed thermodynamic state to the \emph{global} recombination rate. Evaluating the $(\Delta,T)$ integral in Eq.~\eqref{eq:Gamma_rec_2D} gives a binned estimate of the hydrogen-only phase-space rate. Applying $\chi_e=1.08$ to account for helium electrons then yields the full-second-moment curve in Figure~\ref{fig:Gamma_rec_Pion_boost}, which closely tracks the direct volume-averaged simulation rate.

Figure~\ref{fig:Gamma_rec_Pion_boost} illustrates why it is useful to rewrite $\langle x^2\rangle_{\Delta,T}=\mathcal{P}_{\rm ion}(\Delta,T,z)\,\langle x\rangle_{\Delta,T}^2$ and interpret $\mathcal{P}_{\rm ion}$ as the \emph{ionization patchiness factor}. The green curve shows the exact \lumina recombination rate. If one replaces $\langle x^2\rangle_{\Delta,T}$ by $\langle x\rangle_{\Delta,T}^2$ everywhere (purple curve), while retaining the physical helium-electron correction $\chi_e=1.08$, the resulting estimate underpredicts recombinations at early times because recombinations weight the \emph{square} of the ionized fraction: within a given $(\Delta,T)$ bin, mixing highly ionized cells with partially ionized or neutral structure increases $\langle x^2\rangle_{\Delta,T}$ relative to $\langle x\rangle_{\Delta,T}^2$. Including $\mathcal{P}_{\rm ion}$ restores this missing enhancement: the model curve built from $\langle x^2\rangle_{\Delta,T}$ (blue) closely tracks the exact global rate, suggesting that $\mathcal{P}_{\rm ion}(\Delta,T,z)$ captures the correct amount of recombination boosting arising from ionization patchiness at fixed overdensity and temperature.

The redshift dependence of this correction is physically intuitive. The ratio panel in Figure~\ref{fig:Gamma_rec_Pion_boost} shows that the exact rate is enhanced by a factor of $1.70$ over the $\langle x\rangle^2$ approximation at early times, when reionization is highly inhomogeneous and many $(\Delta,T)$ bins contain a broad mixture of ionization states. As reionization proceeds and the ionization field within a given thermodynamic state becomes more uniform, the boost diminishes and approaches unity near overlap.

\begin{figure*}
    \centering
    \safeincludegraphics[width=\textwidth]{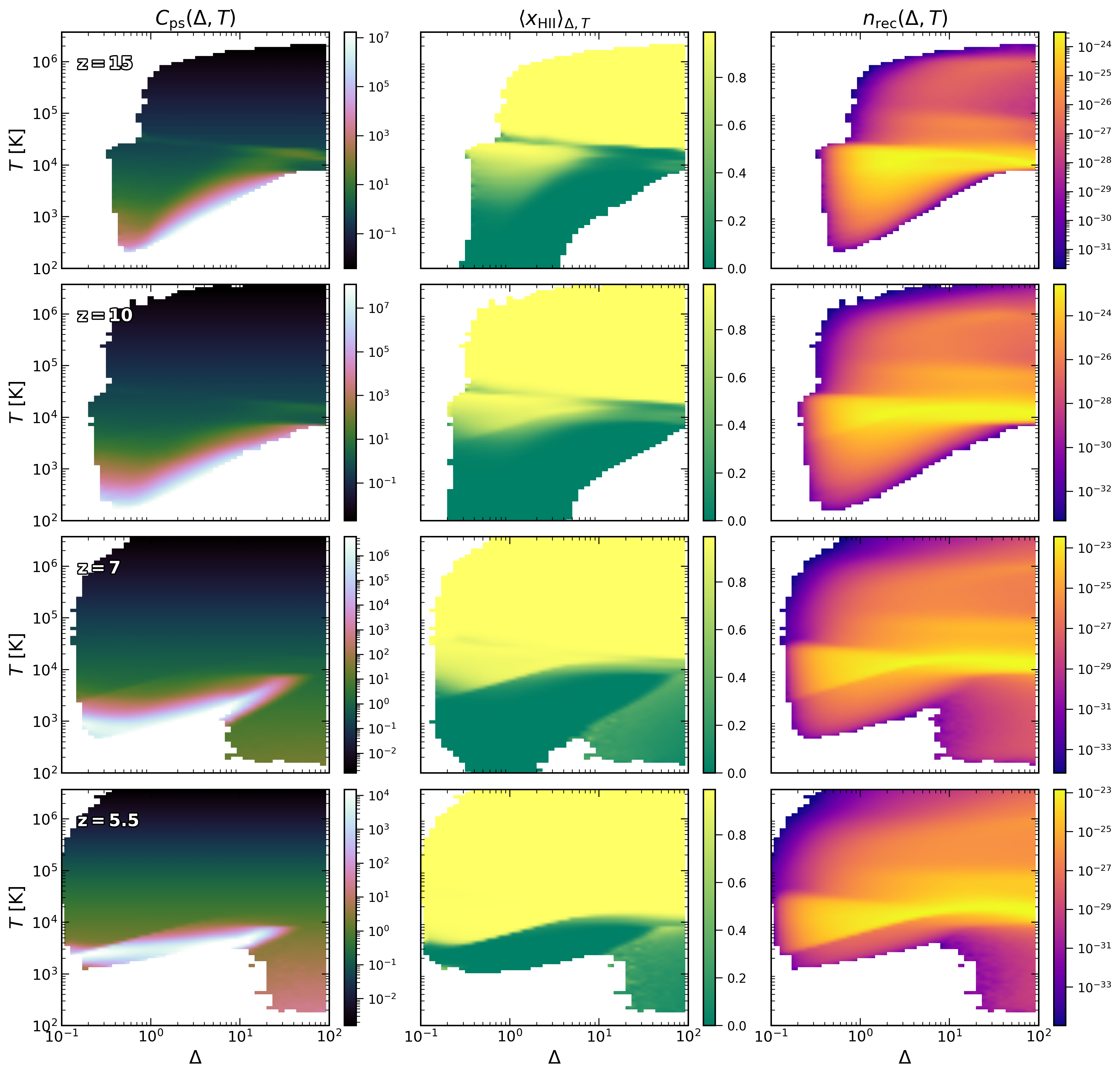}
    \caption[Phase-space clumping and recombination accounting in LUMINA.]{
    Phase-space clumping and recombination accounting in \lumina at four representative redshifts.
    Each row corresponds to a snapshot (redshift shown in the top-left corner of the left panel).
    \textit{Left:} phase-space clumping factor $\mathcal{C}_{\rm ps}(\Delta,T)$ (Eq.~\ref{eq:Cloc_final}), which encodes the intrinsic enhancement of recombinations at fixed $(\Delta,T)$ due to ionization patchiness and the temperature dependence of $\alpha_B(T)$.
    \textit{Middle:} conditional mean ionized fraction $\langle x_{\HII}\rangle_{\Delta,T}$ computed from the 3D volume histogram $V(\Delta,T,x_{\HII})$.
    \textit{Right:} recombination contribution density $n_{\rm rec}(\Delta,T)$ (Eq.~\ref{eq:nrec_phase}), proportional to $\mathcal{C}_{\rm ps}\,\Delta^2\,\langle x_{\HII}\rangle_{\Delta,T}^2\,P_V(\Delta,T)$, whose integral over $(\Delta,T)$ yields the global recombination rate $\Gamma_{\rm rec}(z)$.
    Large $\mathcal{C}_{\rm ps}$ values occur predominantly in cool gas ($T\sim10^3$--$10^4\,\mathrm{K}$), indicating strong ionization patchiness in that regime, while gas above $T\sim10^4\,\mathrm{K}$ is nearly fully ionized and smooth.
    Despite the persistence of highly clumped pockets at late times, their contribution to recombinations becomes small because they occupy little volume and/or have low $\langle x_{\HII}\rangle_{\Delta,T}$, so $n_{\rm rec}$ remains dominated by the photoheated band near $10^4\,\mathrm{K}$.
    }
    \label{fig:stacked_maps_ps}
\end{figure*}

Figure~\ref{fig:stacked_maps_ps} summarizes the physical content of our phase-space decomposition by showing, at four representative redshifts, three $(\Delta,T)$ maps constructed directly from the 3D histogram $V(\Delta,T,x_{\HII})$ in \lumina.  The left column shows the phase-space clumping factor $\mathcal{C}_{\rm ps}(\Delta,T)$ (Eq.~\ref{eq:Cloc_final}), the middle column shows the conditional mean ionized fraction $\langle x_{\HII}\rangle_{\Delta,T}$, and the right column shows the recombination contribution density $n_{\rm rec}(\Delta,T)$ (Eq.~\ref{eq:nrec_phase}).  By construction, integrating the right-hand map over phase space reproduces the global recombination rate, $\Gamma_{\rm rec}(z)=\int \text{d}\Delta\,\text{d}T\;n_{\rm rec}(\Delta,T,z)$.  The three maps therefore separate the problem into three conceptually distinct ingredients: (\textit{i}) the intrinsic enhancement from ionization patchiness and temperature, encoded in $\mathcal{C}_{\rm ps}$; (\textit{ii}) the typical ionization state at fixed $(\Delta,T)$, encoded in $\langle x_{\HII}\rangle_{\Delta,T}$; and (\textit{iii}) the net recombination weight once phase-space occupancy is included through $P_V(\Delta,T)$.

The most immediate feature is that large values of $\mathcal{C}_{\rm ps}$ are confined to relatively cool gas, $T\sim 10^{3}$--$10^{4}\,\mathrm{K}$, while the hot photoheated phase at $T\gtrsim 10^{4}\,\mathrm{K}$ is nearly uniform in ionization, with $\mathcal{P}_{\rm ion}\approx 1$, whereas $\mathcal{C}_{\rm ps}\approx\alpha_B(T)/\alpha_4$ retains the thermal weighting.  This indicates that the dominant source of phase-space clumping is \emph{ionization patchiness} in cool gas instead of generic density structure at fixed overdensity. Within a narrow thermodynamic state, the ionization field contains a mixture of highly ionized and weakly ionized cells, boosting $\langle x^2\rangle_{\Delta,T}$ relative to $\langle x\rangle_{\Delta,T}^2$.  Once gas is strongly photoheated and has had time to thermally relax, the temperature dependence of $\alpha_B(T)$ reduces the recombination efficiency of the ionized phase, and the corresponding enhancement becomes small.

The middle column shows that the ionization state is strongly temperature-selected.  At all redshifts, gas above $T\sim 10^{4}\,\mathrm{K}$ is almost fully ionized, $\langle x_{\HII}\rangle_{\Delta,T}\simeq 1$, with only weak penetration to lower temperatures.  Below $10^{4}\,\mathrm{K}$ the ionized fraction drops rapidly, and the transition becomes increasingly sharp toward later times.  In this sense, temperature is a strong statistical discriminator between the highly ionized, thermally relaxed phase and the partially ionized phase where ionization fronts, shielding, and recent heating produce a broad distribution in $x_{\HII}$.

The right column shows that $n_{\rm rec}(\Delta,T)$ inherits the same phase-space structure previously identified in $P_{\rm rec}(\Delta,T)$: recombinations concentrate into a narrow band around $T\sim 10^{4}\,\mathrm{K}$, with a pronounced low-temperature extension at early times. This similarity is expected, since both quantities trace the same integrand, up to normalization. The key point, however, is that the regions with the largest intrinsic enhancement $\mathcal{C}_{\rm ps}$ are \emph{not} automatically the regions that dominate the recombination budget. Whether phase-space clumping matters globally is controlled by the additional weights in Eq.~\eqref{eq:nrec_phase}, particularly the occupied volume fraction $P_V(\Delta,T)$ and the squared mean ionization state $\langle x_{\HII}\rangle_{\Delta,T}^2$.

This explains the redshift evolution. At high redshift, when reionization is incomplete and the ionization field is highly inhomogeneous, gas elements sharing a $(\Delta,T)$ bin can have different ionization states and hence large $\mathcal{C}_{\rm ps}$. The resulting recombination enhancement depends on the density, ionization, and volume weights of those bins, and does not require a large volume of gas with intermediate ionization fractions. At later times, large $\mathcal{C}_{\rm ps}$ values persist in pockets of cool gas, but these regions occupy little volume and are typically weakly ionized; consequently the factor $\langle x_{\HII}\rangle_{\Delta,T}^2 P_V(\Delta,T)$ suppresses their contribution and recombinations become dominated by the photoheated band near $10^{4}\,\mathrm{K}$. The large local phase-space and ionization-patchiness boosts therefore become less important to the global recombination budget as they are confined to regions that carry little recombination weight. The global density and recombination clumping factors can nevertheless remain significant or grow toward later times.

Taken together, Figure~\ref{fig:stacked_maps_ps} provides a compact accounting identity for recombinations in the simulations. The map $\mathcal{C}_{\rm ps}(\Delta,T)$ isolates where ionization patchiness and temperature can in principle enhance recombinations, while $\langle x_{\HII}\rangle_{\Delta,T}$ and $P_V(\Delta,T)$ determine whether that enhancement is actually realized in the global budget. This separation makes a phase-space clumping description useful for sub-grid and semi-analytic modeling because the state-dependent recombination boost contributes little in regions that are fully ionized and smooth or too rare or neutral to matter.

\section{Discussion}\label{sec:discussion}

\textit{Density clumping versus thermal regulation:}
We find that the recombination-weighted clumping factor $\mathcal{C}_\mathrm{rec}$ is systematically smaller than the standard ionized-hydrogen clumping factor $\mathcal{C}_\mathrm{\HII}$ across all simulations (Figures~\ref{fig:clumping-compare}--\ref{fig:Nrec-compare}). The difference increases toward lower redshift and reflects the thermal regulation of recombinations: once gas is photoheated to temperatures near or above $10^4\,\mathrm{K}$, the hydrogen recombination coefficient decreases substantially, as described by HG97 and \citet{McQuinn2016}. Thus, gas that appears highly clumped in density does not necessarily recombine as efficiently as a density-only model would predict. In \lumina, this produces instantaneous overprediction factors of $1.34$ at $z\simeq8$ and $2.02$ at $z\simeq5$, and a cumulative overprediction factor of $1.53$ by $z\simeq5$ (Figure~\ref{fig:photon_budget}). This supports earlier studies showing that photoheating lowers the effective recombination rate in the ionized IGM through the temperature dependence of $\alpha_B(T)$ \citep[e.g.][]{Pawlik2009,Finlator2012,Daloisio2018}.

This has direct implications for homogeneous reionization models. Prescriptions based only on $\mathcal{C}_\mathrm{\HII}$ tend to overestimate the ionizing photon budget required to maintain ionization, especially after the IGM and CGM have been heated. A more physical reduced model should therefore track not only density inhomogeneity, but also the thermal state of the recombining gas. Classical ionization-balance models usually compress unresolved recombinations into a single clumping factor \citep[][]{Madau1999,Furlanetto2004,Chen2020SCORCH}. However, previous work has shown that this factor depends on resolution, density cuts, ionization state, photoheating, and the treatment of absorbers \citep[e.g.][]{Pawlik2009,RaicevicTheuns2011,Finlator2012,JeesonDaniel2014,KaurovGnedin2015,Sobacchi2014,Mao2020,Bianco2021}. Our analysis extends this picture by separating density structure, temperature dependence, ionization patchiness, and phase-space occupancy within the same radiation-hydrodynamic framework.

A useful comparison point is the recent literature that infers effective clumping factors from the ionizing mean free path and photoionization rate near the end of reionization \citep{Davies2024}. The effective factor in that approach parameterizes the absorption, or equivalently ionization, rate of photons after they escape galaxies. As such, it does not directly measure a recombination rate or impose a recombination-density threshold. Connecting it to a recombination-based clumping factor requires assuming photoionization equilibrium and adopting a density threshold. The comparison is therefore not one-to-one. Our $\mathcal{C}_{\rm rec}$ is instead measured directly from cell-based recombination rates using a $\Delta<100$ numerator, all-gas denominator means, and simulation-consistent Case~B temperature weighting. We therefore interpret these quantities as complementary rather than contradictory, and emphasize that comparisons between ``clumping factors'' require the gas definition and weighting scheme to be stated explicitly.

\textit{Ionized fraction as a natural clock:}
When the clumping factors are reparametrized as functions of the global ionized fraction $x_\mathrm{\HII}$ rather than redshift, the results from \thesanone, \thesantwo, and \lumina become much more similar (Figure~\ref{fig:clumping-compare}). This suggests that the buildup of recombination structure is more closely tied to the ionization state of the IGM than to cosmic time alone. Similar conclusions have appeared in studies of reionization morphology and ionization-fraction-dependent fitting, which found that several large-scale properties of ionized regions are more tightly connected to the global ionized fraction than to redshift or source details \citep[e.g.][]{McQuinn2007,Furlanetto2004,Mesinger2011,Chen2020SCORCH,Wells2024}.

Resolution effects and model differences remain visible. At fixed $x_\mathrm{\HII}$, the higher-resolution \thesanone simulation generally exhibits slightly enhanced clumping compared to lower-resolution runs, consistent with improved resolution of dense filaments and halo outskirts. This is expected because recombinations scale quadratically with density. Nevertheless, the general similarity of the $\mathcal{C}(x_\mathrm{\HII})$ relations suggests that the global ionized fraction may provide a more useful variable for calibrating recombination corrections than the redshift alone, as long as the residual scatter is retained in the uncertainty budget.

\textit{Spatial localization of recombination sinks:}
Recombinations are highly localized throughout reionization. The overdensity PDFs, Lorenz curves, Gini coefficients, and zone fractions show that a small fraction of the selected IGM volume dominates the recombination budget (Figures~\ref{fig:pdf-z6}, \ref{fig:lorenz-allz}, and \ref{fig:zones}). At $z\simeq6$, the densest $\sim10\%$ of the IGM volume accounts for $\gtrsim80\%$ of recombinations, while gas with $\Delta\gtrsim10$ occupies well below $1\%$ of the volume but contributes a large fraction of the total rate. This indicates that recombinations act as clustered photon sinks associated with dense filaments, halo outskirts, and partially self-shielded systems rather than as a uniform loss term in the diffuse IGM.

This strong localization is consistent with analytic and numerical models in which minihalo gas, halo outskirts, and Lyman-limit systems regulate bubble growth, the ionizing mean free path, and the late stages of reionization \citep[e.g.][]{MiraldaEscude2000,FurlanettoOh2005,McQuinn2011,Rahmati2013,Sobacchi2014,Munoz2016Sinks,Chan2024Minihaloes,TheunsChan2024LLS,DaviesFurlanetto2016,lewis2022short}. Any homogeneous model that neglects this localization must absorb it into an effective photon budget or an effective clumping factor.

\textit{Thermal phase structure of recombinations:}
Mapping recombinations into $(\Delta,T)$ phase space shows that the dominant recombination contribution lies along a narrow thermal ridge near $T\sim10^4\,\mathrm{K}$ (Figures~\ref{fig:heatmaps} and \ref{fig:zones}). This ridge is well described by simple equilibrium curves balancing photoheating, recombination cooling, collisional cooling, adiabatic cooling, and Compton cooling. Its low-density width reflects the thermal memory of reionization and expansion cooling, while its high-density extension is shaped by self-shielding and reduced local photoionization rates.

The appearance of this ridge connects recombination sinks to the broader theory of the temperature--density relation of the photoionized IGM developed by HG97 and subsequent work \citep{Theuns1998,Schaye1999,Ricotti2000,McQuinn2016}. It also supports the idea that recombination corrections can be improved by using effective temperatures or phase-space thermal weights rather than assuming a fixed $10^4\,\mathrm{K}$ recombination coefficient.

\textit{Phase-space clumping and ionization patchiness:}
We introduced a phase-space clumping factor $\mathcal{C}_{\rm ps}(\Delta,T,z)$ that separates the intrinsic recombination boost at fixed thermodynamic state from the amount of gas occupying that state. This definition differs from spatially-local clumping factors, which measure density or ionization structure inside real-space subvolumes \citep[e.g.][]{KaurovGnedin2015,Sobacchi2014,Mao2020,Bianco2021}. Instead, $\mathcal{C}_{\rm ps}$ measures a phase-space-local enhancement: it captures how ionization patchiness and temperature modify recombinations among gas cells with the same overdensity and temperature.

The comparison between the exact recombination rate and the estimate obtained by replacing $\langle x_{\HII}^2\rangle_{\Delta,T}$ with $\langle x_{\HII}\rangle_{\Delta,T}^2$ shows that ionization patchiness can produce a non-negligible boost (Figure~\ref{fig:Gamma_rec_Pion_boost}). This correction enhances the rate by a factor of $1.70$ early in reionization, when ionization fronts create a broad mixture of neutral, partially ionized, and ionized cells within the same $(\Delta,T)$ bins, and approaches unity near overlap. In this sense, $\mathcal{P}_{\rm ion}$ isolates ionization inhomogeneity, while $\mathcal{C}_{\rm ps}$ folds in the temperature weighting through $\alpha_B(T)/\alpha_4$.

This behavior follows even for a strictly binary ionization field within a fixed $(\Delta,T)$ bin: if a fraction $p$ of the bin volume is ionized with $x_{\HII}=1$ and the remainder is neutral, then $\langle x_{\HII}\rangle=\langle x_{\HII}^2\rangle=p$ and hence $\mathcal{P}_{\rm ion}=1/p$. Thus, the large early-time values show that overdensity and temperature alone do not completely separate neutral from ionized gas; cells in the same bin can have different proximity to sources, radiation intensity, ionization-front passage, shock histories, and thermal histories. A direct native-cell decomposition at $z\simeq15$ gives a total patchiness boost of $1.70$. Of the excess above the squared-mean approximation, $65\%$ arises at $3<\Delta<30$, $27\%$ at $30<\Delta<100$, $68\%$ at $10^4<T<2\times10^4\,{\rm K}$, and $32\%$ at $10^3<T<10^4\,{\rm K}$. As ionized regions fill the volume, the ionization state becomes more uniform within each bin and the boost approaches unity.

The phase-space maps also show why the largest intrinsic boosts are not necessarily the dominant global sinks. Large $\mathcal{C}_{\rm ps}$ values occur mainly in cool gas, $T\sim10^3$--$10^4\,\mathrm{K}$, where different ionization states coexist. However, much of this gas is weakly ionized or occupies little volume at late times, so its contribution is suppressed by $\langle x_{\HII}\rangle_{\Delta,T}^2$ and $P_V(\Delta,T)$. The global recombination rate remains dominated by the photoheated $T\sim10^4\,\mathrm{K}$ ridge, where gas is both sufficiently ionized and sufficiently abundant.

\textit{Zone-based interpretation:}
The zone-based decomposition in $(\Delta,T)$ space provides a compact physical picture of where ionizing photons are consumed (Figures~\ref{fig:zones} and \ref{fig:zone_evolution}). Early in reionization, recombinations receive substantial contributions from diffuse PI~IGM gas. As reionization proceeds, recombination activity migrates toward higher overdensities: SS~OD peaks at intermediate redshift, while PI~OD and HTNE become increasingly important later. At $z\simeq6$ in \lumina, PI~IGM remains the largest single contributor, but the tiny combined volume occupied by PI~OD and SS~OD supplies nearly half of all recombinations.

The sum of the zone-resolved contributions reproduces the total recombination rate by construction, showing that the decomposition captures the full recombination budget. It therefore provides a physically interpretable way to translate detailed radiation-hydrodynamic outputs into simplified photon-sink categories.

Several limitations should be kept in mind. Our main analysis applies the conventional cut $\Delta<100$ to the recombination and second-moment numerators to exclude dense ISM gas while retaining IGM and CGM recombination sinks. As shown explicitly in Appendix~\ref{app:threshold_sensitivity}, the absolute recombination normalization is sensitive to this choice: increasing $\Delta_{\max}$ adds dense CGM or near-ISM gas whose contribution is amplified by the explicit $\Delta^2$ weighting in the recombination integral. Thus, the fiducial $\Delta<100$ numerator selection should be interpreted as a conservative IGM/CGM sink definition rather than as an exhaustive accounting of all recombinations in dense halo gas. However, the threshold study also shows that the main physical conclusions are not artifacts of this particular cut. The separation between fixed-temperature and temperature-dependent recombination estimates persists over the range of thresholds considered, and the phase-space interpretation of the dominant recombination ridge remains qualitatively unchanged.

The analytic thermal bands used here are equilibrium guides rather than full non-equilibrium thermal histories, and different source spectra or radiation-transport methods could shift the detailed location of the recombination ridge. Finally, while \lumina provides a much larger cosmological volume, its lower resolution relative to \thesanone means that the smallest recombination sinks remain only partially resolved. These limitations mainly affect the detailed normalization of the sink budget; the broader result that recombinations are density-weighted, temperature-regulated, and concentrated in specific regions of phase space is expected to be more robust.

\section{Conclusions}
\label{sec:conclusions}

We have used the \thesanone, \thesantwo, and \lumina radiation-hydrodynamic simulations to study hydrogen recombination sinks during cosmic reionization, with a focus on how density, temperature, ionization state, and phase-space occupancy enter reduced clumping prescriptions. Our main conclusions are:
\begin{enumerate}[leftmargin=*]
  \item The recombination-weighted clumping factor $\mathcal{C}_{\rm rec}$ is systematically smaller than the standard density-only ionized-hydrogen clumping factor $\mathcal{C}_{\rm \HII}$. In \lumina, the density-only prescription overpredicts the instantaneous recombination rate by factors of $1.34$ at $z\simeq8$ and $2.02$ at $z\simeq5$, and the cumulative count by a factor of $1.53$ by $z\simeq5$. The growing offset shows that temperature-dependent recombination physics cannot be captured by a fixed-$10^4\,\mathrm{K}$ density-only prescription.

  \item When expressed as functions of the volume-weighted global ionized fraction $x_{\HII}$, the clumping histories of \thesanone, \thesantwo, and \lumina follow an approximately common relation at the $\sim10$--$20\%$ level. This makes $x_{\HII}$ a more informative variable for clumping than redshift alone, although the relation is not strictly universal and absolute recombination rates still depend on how each simulation samples the density field at a given ionization state.

  \item Recombinations are highly localized in the selected IGM/CGM gas. At $z\simeq6$, a small fraction of the volume contains most of the recombination budget, indicating that dense filaments, halo outskirts, and partially self-shielded gas act as clustered photon sinks rather than as a spatially uniform loss term.

  \item In $(\Delta,T)$ phase space, the dominant recombination contribution follows a narrow thermal ridge near $T\sim10^4\,\mathrm{K}$. Simple equilibrium guide curves provide a useful interpretation of this structure and offer a compact connection between IGM/CGM thermodynamics and reduced recombination models.

  \item The phase-space clumping factor $\mathcal{C}_{\rm ps}(\Delta,T,z)$ isolates the intrinsic boost from ionization patchiness and temperature at fixed thermodynamic state. The patchiness boost enhances the recombination rate by a factor of $1.70$ early in reionization and approaches unity near overlap. Its global importance, however, is set by additional weights from density, ionized fraction, and occupied volume. For this reason, the contribution-weighted quantity $n_{\rm rec}(\Delta,T)$ is the most direct diagnostic for global recombination accounting.
\end{enumerate}

More broadly, our results suggest that reduced sink prescriptions should distinguish between density structure, temperature weighting, and ionization patchiness rather than compressing all three effects into a single redshift-dependent clumping factor. In future work it would be valuable to compare these calibrations across other radiation-hydrodynamic simulations, including CROC \citep{Gnedin2014design,croc120cMpc}, CoDa \citep{ocvirk2016cosmic,ocvirk2020cosmic}, SPHINX \citep{rosdahl2018sphinx,sphinx2022}, Aurora \citep{aurora2017}, the semi-numerical AMBER model \citep{Trac_2022}, SPICE \citep{bhagwat2024spice}, SAGUARO \citep{Cain2026SAGUARO}, \thesanzoom \citep{Kannan2025}, and related suites, as well as observationally inferred effective sink budgets. This could further inform how much of the remaining scatter is physical and how much is definitional, e.g., feedback prescriptions, radiative-transfer methods, and the treatment of Jeans-scale IGM structure. In parallel, a forthcoming study by Bian et al. (in prep) will analyze ionizing escape fractions in \lumina, providing a complementary measurement of the source term in the reionization photon budget. Combined with the sink terms quantified in this work, these results will enable direct tests of both simple ionization-balance ODE models and more sophisticated semi-analytic or semi-numerical reionization frameworks \citep[e.g.][]{Madau1999,Furlanetto2004,Mesinger2011,Chen2020SCORCH,Gnedin2022Review}.

\section*{Acknowledgments}
We thank the referee for constructive comments and suggestions, which have improved the quality of this work.
An award of computer time was provided by the INCITE program. This research used resources of the Oak Ridge Leadership Computing Facility at the Oak Ridge National Laboratory, which is supported by the Advanced Scientific Computing Research programs in the Office of Science of the U.S. Department of Energy under Contract No.\ DE-AC05-00OR22725.
The authors acknowledge the MIT Office of Research Computing and Data, FAS Division of Science Research Computing Group at Harvard University, and High Performance Computing at The University of Texas at Dallas (HPC@UTD) for providing resources that have contributed to the research results reported within this paper.
Support for programs JWST-AR-08709 (AS) and JWST-AR-04814 (XS, MV) were provided by NASA through a grant from the Space Telescope Science Institute, which is operated by the Association of Universities for Research in Astronomy, Inc., under NASA contract NAS 5-03127.
Support for OZ was provided by Harvard University through the Institute for Theory and Computation Fellowship.
RK acknowledges support of the Natural Sciences and Engineering Research Council of Canada (NSERC) through a Discovery Grant and a Discovery Launch Supplement (funding reference numbers RGPIN-2024-06222 and DGECR-2024-00144) and York University's Global Research Excellence Initiative.
MV acknowledges support through NASA ATP Grant 23-ATP23-149 and NSF AAG Grant AST-2307699.
This work was supported by the LOEWE programme of the State of Hesse (LOEWE-Spitzen-Professur).
VS and LH acknowledge support from the Simons Foundation through the ``Learning the Universe'' initiative.

\bibliographystyle{mnras}
\bibliography{main}

\appendix
\setcounter{equation}{0}
\renewcommand{\theequation}{\thesection\arabic{equation}}
\renewcommand{\theHequation}{appendix.\thesection.\arabic{equation}}

\section{Derivation of Thermal Equilibrium Bands}
\label{app:thermal_bands}

This appendix derives the thermal-equilibrium bands summarized in Section~\ref{sec:thermal_bands} and used throughout the phase-space analysis. Following HG97 and related treatments \citep{Theuns1998,TracCenLoeb2008}, we start from the thermal evolution of an ideal monatomic plasma in terms of the internal energy density $u$,
\begin{equation}
\frac{{\rm d}u}{{\rm d}t}
=
\mathcal{H}_{\rm ph}
-\mathcal{L}_{\rm rad}
-\mathcal{L}_{\rm ad}
-\mathcal{L}_{\rm Comp}
+\mathcal{H}_{\rm struct} \, ,
\label{eq:dudt_general_full}
\end{equation}
where $\mathcal{H}_{\rm ph}$ denotes photoheating by ionizing radiation, $\mathcal{L}_{\rm rad}$ radiative cooling, $\mathcal{L}_{\rm ad}$ adiabatic cooling from cosmic expansion, $\mathcal{L}_{\rm Comp}$ Compton cooling from interactions with CMB photons, and $\mathcal{H}_{\rm struct}$ heating associated with shocks, gravitational collapse, and other structure-formation processes. For the diffuse photoionized IGM considered here, we neglect explicit structure-formation heating and set $\mathcal{H}_{\rm struct}\simeq0$.

\textit{Heating and cooling ingredients:}
The hydrogen-dominated volumetric terms entering Eq.~\eqref{eq:dudt_general_full} are
\begin{align}
  \mathcal{H}_{\rm ph} &= n_{\HI}\,\Gamma_{\HI}\,\langle E_{\rm heat}\rangle \, , \label{eq:Hph_vol}\\[4pt]
  \mathcal{L}_{\rm rad} &= n_e n_{\HII}\,\Lambda_{\rm rec}^{\HII}(T)
  + n_e n_{\HI}\,\Lambda_{\rm exc}^{\HI}(T) \, , \label{eq:Lrad_vol}\\[4pt]
  \mathcal{L}_{\rm ad} &= 3\,k_{\rm B}\,T\,n_{\rm tot}\,H \, , \label{eq:Lad_vol}\\[4pt]
  \mathcal{L}_{\rm Comp} &=
  \frac{3}{2}\,k_{\rm B}\,n_e\,K_{\rm C}(z)\,
  (T-T_{\rm CMB}) \, . \label{eq:LComp_vol}
\end{align}
Here $\langle E_{\rm heat}\rangle$ denotes the mean photoheating per ionization, $T_{\rm CMB}(z)=2.7255(1+z)\,{\rm K}$, and $\Lambda_{\rm rec}^{\HII}(T)$ and $\Lambda_{\rm exc}^{\HI}(T)$ describe recombination and collisional-excitation cooling, respectively. The Compton coupling coefficient is
\begin{equation}
K_{\rm C}(z)=
\frac{8\sigma_{\rm T}a_{\rm rad}T_{\rm CMB}^4}{3m_e c} \, .
\label{eq:KC_def}
\end{equation}

\textit{Photoionization--recombination equilibrium:}
In the highly ionized limit, $x_{\HI}\ll1$, the neutral hydrogen abundance is set approximately by
\begin{equation}
  n_{\HI}\,\Gamma_{\HI}
  \simeq
  n_e n_{\HII}\,\alpha_{\rm B}(T) \, ,
  \label{eq:photo-recomb balance}
\end{equation}
which gives
\begin{equation}
  x_{\HI}(T,\Delta)
  \simeq
  \frac{\alpha_{\rm B}(T)\,n_{\rm H}(\Delta,z)}
       {\Gamma_{\HI}} \, .
\label{eq:photoion_eq_full}
\end{equation}
This relation supplies the density dependence of the collisional-excitation term.

\textit{Pair-normalized rates:}
Because the dominant microphysical heating and cooling terms scale with the number of interacting electron--proton pairs, we divide Eq.~\eqref{eq:dudt_general_full} by $n_e n_{\HII}$. Using Eq.~\eqref{eq:photo-recomb balance}, $n_{\rm tot}=n_{\rm H}(1+f_{\rm He}+x_e)$, with $f_{\rm He}\equiv Y/(4X)$ and $\mu^{-1}=X(1+x_e)+Y/4$, taking $X=0.76$ and $Y=0.24$, and approximating $n_e n_{\HII}\simeq n_{\rm H}^2$ in the highly ionized limit, we obtain
\begin{align}
  \alpha_{\rm B}(T)\,\langle E_{\rm heat}\rangle
  &\quad \text{\footnotesize (photoheating)} \, ,
  \label{eq:heat_pair}\\
  \Lambda_{\rm rec}^{\HII}(T)
  &\quad \text{\footnotesize (recombination cooling)} \, ,
  \label{eq:Lrec_pair}\\
  x_{\HI}(T,\Delta)\,\Lambda_{\rm exc}^{\HI}(T)
  &\quad \text{\footnotesize (collisional-excitation cooling)} \, ,
  \label{eq:Lexc_pair}\\
  \frac{3k_{\rm B}H(z)T}
  {X\mu\,n_{\rm H}(\Delta,z)}
  &\quad \text{\footnotesize (adiabatic cooling)} \, ,
  \label{eq:Lad_pair}\\
  \frac{3k_{\rm B}K_{\rm C}(z)f_e}
  {2X\mu\,n_{\rm H}(\Delta,z)}
  (T-T_{\rm CMB})
  &\quad \text{\footnotesize (Compton cooling)} \, .
  \label{eq:LComp_pair}
\end{align}
Here $f_e\equiv x_e/(1+f_{\rm He}+x_e)$, $n_{\rm H}(\Delta,z)=X\rho_b(z)\Delta/m_\text{H}$, and $\rho_b(z)=\rho_{\rm crit,0}\Omega_b(1+z)^3$. We adopt the hydrogen recombination- and collisional-excitation-cooling rates and rate coefficients from standard references \citep{Cen1992,VernerFerland1996} and HG97.

\textit{Equilibrium-band construction:}
Setting ${\rm d}u/{\rm d}t=0$ and balancing all of the pair-normalized heating and cooling terms gives the final lower photoheated solution $T_{\rm low}(\Delta|z)$:
\begin{align}
  &\alpha_{\rm B}(T)\,\langle E_{\rm heat}\rangle
  =
  \Lambda_{\rm rec}^{\HII}(T)
  + x_{\HI}(T,\Delta)\,\Lambda_{\rm exc}^{\HI}(T)
  \notag\\
  &\quad+
  \frac{3k_{\rm B}H(z)T}
  {X\mu\,n_{\rm H}(\Delta,z)}
  +
  \frac{3k_{\rm B}K_{\rm C}(z)f_e}
  {2X\mu\,n_{\rm H}(\Delta,z)}
  (T-T_{\rm CMB}) \, .
  \label{eq:app_Tlow_implicit_full}
\end{align}
Dropping the adiabatic and Compton terms gives the final upper solution $T_{\rm high}(\Delta|z)$:
\begin{equation}
  \alpha_{\rm B}(T)\,\langle E_{\rm heat}\rangle
  =
  \Lambda_{\rm rec}^{\HII}(T)
  + x_{\HI}(T,\Delta)\,\Lambda_{\rm exc}^{\HI}(T) \, .
  \label{eq:app_Thigh_implicit_full}
\end{equation}
These are the fully derived forms of the equilibrium-band equations summarized and used in the main text as Eqs.~\eqref{eq:Tlow_implicit_full} and \eqref{eq:Thigh_implicit_full}. We use fixed effective photoheating energies $\langle E_{\rm heat}\rangle=3.25\,{\rm eV}$ for \thesanone and \thesantwo and $4.20\,{\rm eV}$ for \lumina, reflecting their different ionizing spectral shapes.

For $T_{\rm low}$ and $T_{\rm high}$ we adopt a fiducial $\Gamma_{\HI}=10^{-12}\,{\rm s^{-1}}$, characteristic of the reionized IGM at $z\sim5$--$6$ with spatial fluctuations \citep[e.g.][]{BoltonHaehnelt2007,FaucherGiguere2009,BeckerBolton2013,HaardtMadau2012,Puchwein2019,DaviesFurlanetto2016}. To model self-shielding, we solve the same balance as $T_{\rm high}$ with reduced local photoionization rates. We define $T_{\rm SS,high}$ using $\Gamma_{\HI}=10^{-13}\,{\rm s^{-1}}$ and $T_{\rm SS,low}$ using $\Gamma_{\HI}=10^{-16}\,{\rm s^{-1}}$, bracketing moderate and strong shielding \citep[cf.][]{Rahmati2013,KaurovGnedin2015}. As $\Gamma_{\HI}$ decreases, these solutions shift toward cooler temperatures at high overdensity and trace the recombination-dominated, self-shielded locus in the $\Delta$--$T$ plane.

\section{Threshold Sensitivity}
\label{app:threshold_sensitivity}

Throughout the main analysis we adopt the fiducial IGM/CGM selection $\Delta<100$. It is expected that the absolute recombination budget depends on this choice: recombinations scale roughly as density squared, and dense gas in halo outskirts and the CGM can contribute significantly even when it occupies a small volume. The analysis does not claim to provide an exhaustive accounting of the full reionization photon budget. The goal of this appendix is to show that the main claims of this paper are not artifacts of the particular threshold $\Delta_{\max}=100$.

The phase-space recombination integral introduced in Section~\ref{sec:local_clumping} provides a convenient way to carry out this threshold test. Consistent with the fiducial analysis, this calculation uses the Case~B recombination coefficient. For any overdensity threshold, we evaluate
\begin{equation}
\begin{aligned}
\Gamma_{\rm rec}(<\Delta_{\max},z)
&= \bar n_{\rm H}^{\,2}
\int^{\Delta_{\max}}\!{\rm d}\Delta
\int {\rm d}T\;
\alpha_B(T)\,\Delta^2 \\
&\quad\times
\langle x_{\HII}^2\rangle_{\Delta,T}
P_V(\Delta,T,z) \, .
\end{aligned}
\label{eq:app_gamma_threshold}
\end{equation}
Changing $\Delta_{\max}$ therefore changes the upper limit of the same physical integral, while leaving the temperature, ionization, and volume weights unchanged.

\begin{figure}
    \centering
    \safeincludegraphics[width=\columnwidth]{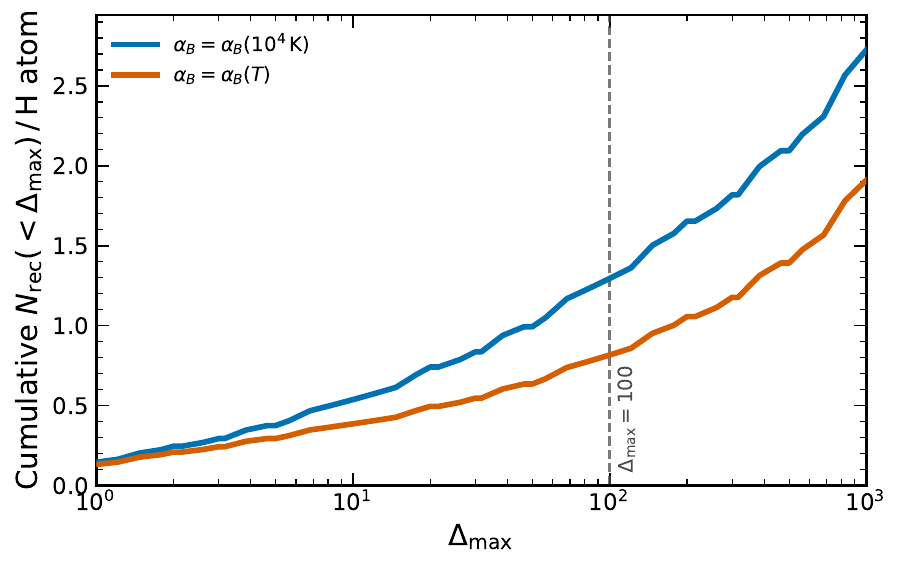}
    \caption{Threshold-sensitivity test for cumulative recombinations per hydrogen atom, integrated to $z\simeq5$ in \lumina, as a function of the overdensity threshold $\Delta_{\max}$ used in the phase-space integral. The blue curve uses a fixed recombination coefficient $\alpha_B(10^4\,{\rm K})$, while the orange curve uses the full temperature-dependent coefficient $\alpha_B(T)$. The vertical dashed line marks the fiducial selection $\Delta_{\max}=100$. At this threshold, the difference between the curves quantifies the effect of temperature weighting within the fiducial IGM/CGM gas selection. The rising normalization with increasing $\Delta_{\max}$ reflects the expected sensitivity of any recombination budget to dense gas, but the persistence of the curve separation suggests that the thermal-suppression conclusion is qualitatively unchanged over the threshold range considered here.}
    \label{fig:app_cumulative_threshold}
\end{figure}

As a first example, Figure~\ref{fig:app_cumulative_threshold} shows the cumulative number of recombinations per hydrogen atom integrated down to $z\simeq5$ as a function of $\Delta_{\max}$. We compare the full temperature-dependent integral with a fixed-temperature calculation in which $\alpha_B(T)$ is replaced everywhere by $\alpha_B(10^4\,{\rm K})$. The total normalization increases as denser gas is included, as expected. However, the separation between the two curves remains present across the full threshold range. At the fiducial value $\Delta_{\max}=100$, the full calculation gives $N_{\rm rec}\simeq0.82$ per hydrogen atom, while the fixed-temperature calculation gives $N_{\rm rec}\simeq1.30$. Thus temperature weighting suppresses the cumulative recombination budget by roughly one third within the fiducial IGM/CGM selection. Similar separations are obtained for other choices of $\Delta_{\max}$, suggesting that the thermal-suppression result is not tied to the specific cut at $\Delta=100$.

We also quantify the instantaneous threshold sensitivity by normalizing the rate at each threshold to the fiducial value,
\begin{equation}
\begin{aligned}
R(\Delta_{\max},z)
&\equiv
\frac{\Gamma_{\rm rec}(<\Delta_{\max},z)}
     {\Gamma_{\rm rec}(<100,z)} \\
&=1+\epsilon(100,\Delta_{\max},z) \, ,
\end{aligned}
\label{eq:app_threshold_ratio}
\end{equation}
where for $\Delta_{\max}>100$
\begin{equation}
\epsilon(100,\Delta_{\max},z)
\equiv
\frac{\Gamma_{\rm rec}(100<\Delta<\Delta_{\max},z)}
     {\Gamma_{\rm rec}(<100,z)} \, .
\label{eq:app_epsilon_def}
\end{equation}
The quantity $\epsilon$ is therefore the fractional extra recombination rate contributed by gas between $\Delta=100$ and the new threshold. For $\Delta_{\max}<100$, the analogous ratio is below unity and measures the fraction of the fiducial recombination rate removed by excluding gas in the upper part of the fiducial IGM/CGM interval.

\begin{figure}
    \centering
    \safeincludegraphics[width=\columnwidth]{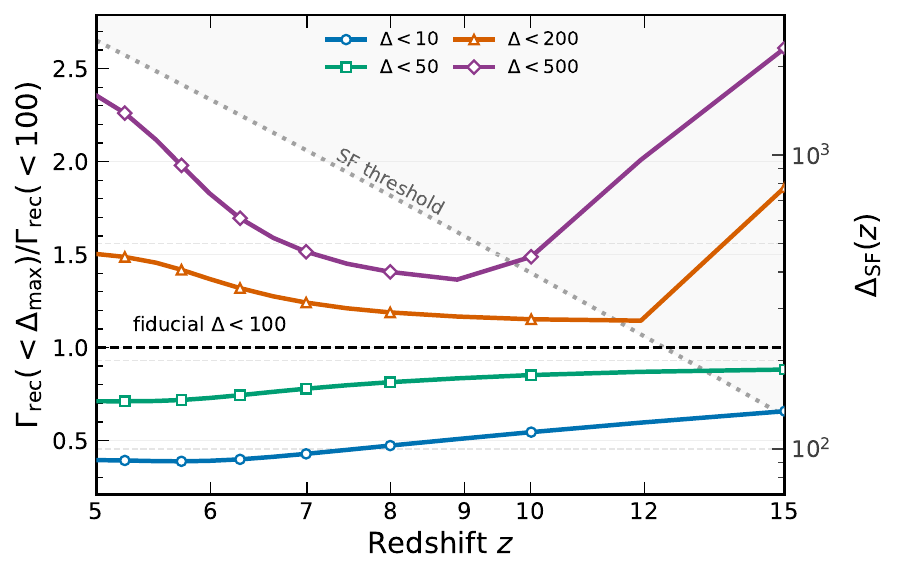}
    \caption{Instantaneous sensitivity of the \lumina recombination rate to the overdensity threshold. The colored curves show $\Gamma_{\rm rec}(<\Delta_{\max})/\Gamma_{\rm rec}(<100)$ for several thresholds, with the dashed horizontal line marking the fiducial normalization. The gray dotted curve, shown on the right axis, gives the overdensity corresponding to the nominal star-formation threshold $n_{\rm H,SF}=0.1\,{\rm cm^{-3}}$. The increasing ratios for $\Delta_{\max}>100$ show that dense gas can add recombination weight, particularly at low redshift, while lower cuts remove a substantial part of the fiducial IGM/CGM sink budget.}
    \label{fig:app_threshold_ratio}
\end{figure}

Figure~\ref{fig:app_threshold_ratio} shows this ratio for representative thresholds. The rate is sensitive to the threshold, especially at later times for the diffuse IGM, because high-density gas becomes increasingly important as reionization progresses; at the highest plotted redshifts ($z\gtrsim12$), the elevated ratios for large $\Delta_{\max}$ are driven by gas at or above the star-formation threshold $\Delta_{\rm SF}(z)$ shown on the secondary axis, which is not part of the diffuse IGM/CGM sink budget. This context illustrates when large values of $\Delta_{\max}$ begin to include gas approaching the star-forming regime. We treat $\Delta<100$ as a conservative IGM/CGM sink definition, not as a claim that gas above $\Delta=100$ is irrelevant.

The direction of this trend follows directly from the structure of Eq.~\eqref{eq:app_gamma_threshold}. The integrand contains the explicit $\Delta^2$ factor, while the high-density volume PDF falls approximately as $P_V\propto\Delta^{-\beta(z)}$ over the fitted high-density tail. In \lumina, we find typical values $\beta\simeq3.79$ at $z\simeq15$ decreasing to $\beta\simeq2.5$ by $z\simeq5$, with a simple fit $\beta(z)\simeq1.82+0.131z$. Thus the high-density volume tail becomes shallower with time: as structure forms, dense regions occupy more volume and therefore carry more recombination weight. This explains why larger choices of $\Delta_{\max}$ increase the recombination normalization most strongly at late times.

This sensitivity does not undermine the main conclusions of the paper. The phase-space decomposition of recombinations depends primarily on the thermodynamic and ionization state of the gas: the dominant ridge near $T\sim10^4\,{\rm K}$, the role of temperature-dependent recombination coefficients, the analytic thermal bands, and the zone definitions are all constructed in $(\Delta,T)$ space and are not tied to the specific value $\Delta_{\max}=100$. Increasing the threshold would reveal additional dense CGM or near-ISM gas with similar phase-space logic, and such gas would be interesting for a focused CGM absorber study. The present work is instead aimed at clumping and recombination sinks in the IGM and diffuse CGM. Within that scope, recombinations are certainly threshold-sensitive in normalization, but the physical claims derived in the main text are qualitatively unchanged under the threshold variations considered here.

\section{Fits of \texorpdfstring{$\mathcal{C}(x_{\HII})$}{C(xHII)}}
\label{app:fits_cxhii}

In Section~\ref{sec:clumping_evolution} we showed that the global clumping factors in \thesanone, \thesantwo, and \lumina follow a much tighter relation when plotted against the global ionized hydrogen fraction $x_{\HII}$ than when plotted against redshift, approximately universal at the $\sim10$--$20\%$ level. Here we provide simple empirical fits to these relations for both $\mathcal{C}_{\rm rec}$ and $\mathcal{C}_{\rm \HII}$. These fits are intended as compact summaries of the simulation trends and may be useful for reduced reionization models that evolve in terms of ionized fraction rather than redshift.

We fit each curve with a power law,
\begin{equation}
\mathcal{C}(x_{\HII}) = A\,x_{\HII}^{m} \, ,
\label{eq:app_cxhii_fit}
\end{equation}
using a linear least-squares fit in $(\log_{10}x_{\HII},\log_{10}\mathcal{C})$ over the simulated ionized-fraction range corresponding to $5.5\le z\le15$ for \thesanone and \thesantwo, and $4.75\le z\le15$ for \lumina. In this parametrization, $A$ is the fitted clumping factor at $x_{\HII}=1$, while $m$ measures how rapidly the clumping factor increases toward earlier, less ionized stages of reionization. Since $x_{\HII}$ decreases toward early times, the fitted values $m<0$ correspond to increasing clumping as the ionized volume fraction decreases. These fits should be regarded as empirical interpolations over the simulated range and should not be extrapolated beyond it.

\begin{table}
\centering
\caption{\justifying \noindent \textup{Power-law fits of the form $\mathcal{C}(x_{\HII})=A\,x_{\HII}^{m}$ for the global clumping factors. The quoted uncertainties are the formal $1\sigma$ uncertainties from the linear fit in log--log space, and the final column gives the rms residual scatter in dex. These uncertainties do not include systematic effects from resolution, box size, the overdensity cut, or the assumed power-law form.}}
\label{tab:app_cxhii_fits}
\footnotesize
\setlength{\tabcolsep}{3.5pt}
\renewcommand{\arraystretch}{1.1}
\begin{tabular}{ccccc}
\toprule
Simulation & Quantity & $A$ & $m$ & rms [dex] \\
\midrule
\thesanone & $\mathcal{C}_{\rm rec}$ &
$3.568\pm0.039$ & $-0.830\pm0.005$ & 0.028 \\
\thesantwo & $\mathcal{C}_{\rm rec}$ &
$2.930\pm0.047$ & $-0.895\pm0.005$ & 0.042 \\
\lumina & $\mathcal{C}_{\rm rec}$ &
$2.811\pm0.038$ & $-0.863\pm0.008$ & 0.016 \\
\midrule
\thesanone & $\mathcal{C}_{\rm \HII}$ &
$4.856\pm0.091$ & $-0.779\pm0.009$ & 0.048 \\
\thesantwo & $\mathcal{C}_{\rm \HII}$ &
$4.103\pm0.091$ & $-0.862\pm0.008$ & 0.057 \\
\lumina & $\mathcal{C}_{\rm \HII}$ &
$4.855\pm0.174$ & $-0.759\pm0.022$ & 0.043 \\
\bottomrule
\end{tabular}
\renewcommand{\arraystretch}{1}
\end{table}

Table~\ref{tab:app_cxhii_fits} quantifies the collapse seen in Figure~\ref{fig:clumping-compare}. The normalizations are all of order a few, showing that once the simulations are highly ionized the residual IGM/CGM clumping is modest. For a fixed simulation, $A$ is larger for $\mathcal{C}_{\rm \HII}$ than for $\mathcal{C}_{\rm rec}$, reflecting the thermal suppression discussed in the main text: a density-only clumping factor assigns too much recombination weight to photoheated gas compared to a temperature-dependent recombination rate.

The fitted slopes are also similar across the three simulations, with $m\simeq -0.76$ to $-0.90$. This is consistent with the interpretation that the growth of clumping toward earlier, less ionized stages is more closely organized by the global ionization state than by redshift itself. The remaining differences are nevertheless physically informative. The parameter $A$ captures the residual late-time clumping level and can depend on resolution, box size, and how ionized gas samples dense environments. The parameter $m$ describes how quickly dense recombination sinks enter the ionized volume as reionization proceeds, and may therefore encode differences in source bias, radiation transport, and the timing of structure growth. In this sense, $A$ and $m$ provide a compact two-parameter characterization of each simulation's clumping history while preserving the approximately common $\mathcal{C}(x_{\HII})$ behavior emphasized in this study.

\end{document}